\documentclass[preprint,notoc]{JHEP3} 
\usepackage{epsfig,multicol}
\usepackage{amssymb}
\usepackage{graphics}

\usepackage{amsmath,latexsym}
\def\beq{\begin{eqnarray}}    
\def\eeq{\end{eqnarray}}      

\newcommand{\OM}{\Omega_M}
\newcommand{\OL}{\Omega_{\Lambda}}

\newcommand{\rc}{\rho_c}
\newcommand{\rM}{\rho_M}
\newcommand{\rR}{\rho_R}
\newcommand{\rL}{\rho_{\Lambda}}
\newcommand{\CC}{\Lambda}
\newcommand{\bCC}{\beta_{\Lambda}}

\title{Running $G$ and $\CC$ at low energies from physics at
$M_X$:\\ possible cosmological and astrophysical implications}
\author{
Ilya L. Shapiro $^{1}$, Joan Sol\`a $^{2}$,
Hrvoje  \v{S}tefan\v{c}i\'{c} $^{3}$\\
$^{1}\,$Departamento de F\'{\i}sica, ICE,
Universidade Federal de Juiz de Fora, Minas Gerais, Brazil,
E-mail: shapiro@fisica.ufjf.br\\
$^{2}\,$Dep. E.C.M.  Universitat de Barcelona, and C.E.R. for
Astrophysics, Particle Physics and Cosmology\,\thanks{Associated
with Instituto de Ciencias del Espacio-CSIC.}, Diagonal 647,
Barcelona, Catalonia,
Spain, E-mail: sola@ifae.es\\
$^{3}$ Theoretical Physics Division, Rudjer Bo\v{s}kovi\'{c}
Institute, P.O. Box 180, HR-10002 Zagreb, Croatia, E-mail:
shrvoje@thphys.irb.hr}

\preprint{ UB-ECM-PF-04/25}

\abstract{The renormalization group (RG) approach to cosmology is
an efficient method to study the possible evolution of the
cosmological parameters from the point of view of quantum field
theory (QFT) in curved space-time. In this work we continue our
previous investigations of the RG method based on potential
low-energy effects induced from physics at very high energy scales
$M_X\lesssim M_P$. In the present instance we assume that both the
Newton constant, $G$, and the cosmological term, $\CC$, can be
functions of a scale parameter $\mu$. It turns out that $G(\mu)$
evolves according to a logarithmic law which may lead to
asymptotic freedom of gravity, similar to the gauge coupling in
QCD. At the same time $\Lambda(\mu)$ evolves quadratically with
$\mu$. We study the consistency and cosmological consequences of
these laws when $\mu\simeq H$. Furthermore, we propose to extend
this method to the astrophysical domain after identifying the
local RG scale at the galactic level. It turns out that Kepler's
third law of celestial mechanics receives quantum corrections that
may help to explain the flat rotation curves of the galaxies
without introducing the dark matter hypothesis. The origin of
these effects (cosmological and astrophysical) could be linked,
in our framework, to physics at $M_X\sim 10^{16-17}\,GeV$.}


\begin{document}

\section{Introduction}
\label{sect:Intro}

The recent developments in physical cosmology have provided an
astonishingly accurate picture of our universe, within the canons
of the Friedmann-Lema\^\i tre-Robertson-Walker (FLRW)
paradigm\,\cite{Peebles}. From these achievements emerged what
has been called the {\em cosmological concordance model},
characterized by an essentially zero value of the spatial
curvature parameter and a non-vanishing positive value of the
cosmological constant (CC) term, $\CC$, in Einstein's equations --
or in general of some form of dark energy (DE) that cannot be
attributed to any form of matter and radiation. This new standard
cosmological model supersedes and refutes (by many standard
deviations) the old Einstein-de Sitter (or critical-density)
cosmological model, which has been the standard cosmological model
until recently. The latter is also a spatially flat model, but it
is characterized by a zero value of the cosmological constant.
Evidence for the values of the cosmological parameters within the
new concordance model comes both from tracing the rate of
expansion of the universe with high-z Type Ia supernovae (SNe Ia)
experiments \cite{R98P99,R04P03} and from the precise measurement
of the anisotropies in the cosmic microwave background (CMB)
radiation\,\cite{WMAP2,WMAP1}. These experiments indicate that the
value of the cosmological constant is around\footnote{The original
CC term, $\lambda$, in Einstein's equations is related to our
$\CC$ by $\lambda=8\pi\,G\,\CC$, where $G$ is Newton's constant.
Notice that $\CC$, in our notation, has dimensions of energy
density, and to emphasize this fact it will sometimes be denoted
$\CC\equiv\rL$.}
\begin{equation}\label{CCvalue}
\CC_0=\OL^0\,\rc^0\simeq 6\,h_0^2\times 10^{-47}\,GeV^{4}\,,
\end{equation}
where $\OL^0\sim 70\%$. {Here $\rc^0 \simeq \left(
3.0\,\sqrt{h_{0}}\times 10^{-12}\,GeV\right) ^{4}$ is the present
value of the critical density and $H_0 = 100\,h_0$ km/s/Mpc, with
$h_{0}= 0.72\pm 0.08$, is the most precisely known value of
Hubble's constant\,\cite{HST03}. In the context of the Standard
Model (SM) of electroweak interactions, the physical (measured)
value (\ref{CCvalue}) should be the sum of the vacuum CC term in
Einstein's equations, $\CC _{vac}$, and the induced contribution
from the vacuum expectation value of the Higgs effective
potential, $\CC_{ind}=\langle V_{\rm eff} \rangle$, namely
$\CC=\CC _{vac}+\CC _{ind}\,.$ Other contributions should be
added within the SM, like those from QCD, but they are much
smaller. It is only the combined parameter $\CC$ that makes
physical sense, whereas both $\CC _{vac}$ and $\CC _{ind}$ remain
individually unobservable-- see e.g. \cite{JHEPCC1} for a more
detailed explanation. From the current LEP numerical bound on the
Higgs boson mass, $\,M_{H}> 114.4\,GeV$ (at the $95\%$ C.L.)
\cite{LEPHWG}, one finds $\,\,\,\left| \CC
_{ind}\right|=(1/8)M_H^2\,v^2> 1.0\times 10^{8}\,GeV^{4}$, where
$v\simeq 246\,GeV$ is the vacuum expectation value of the Higgs
field. Clearly, $\left| \CC _{ind}\right|$ is $55$ orders of
magnitude larger than the observed CC value (\ref{CCvalue}). Such
discrepancy, the so-called ``old'' CC problem
\cite{weinRMP,CCRev}, manifests itself in the necessity of
enforcing an unnaturally exact fine tuning of the original
cosmological term $\CC _{vac}$ in the vacuum action that has to
cancel the induced counterpart $\,\CC _{ind}$ within a precision
(in the SM) of one part in $\,10^{55}$. But the CC problem seems
to be quite more general and it seems to plague all physical
theories (QFT and string theory) making use of the concept of
vacuum energy\, \cite{weinRMP,CCRev,wittenDM}. Moreover, all
attempts to deduce the small value of the cosmological constant
from a sound theoretical idea involve some explicit or implicit
form of severe fine-tuning among the parameters of the model. This
became clear already from the first attempts to treat the dark
energy component as a dynamical scalar field
\cite{Dolgov,PSW,Aguila}. More recently this approach took the
popular form of a ``quintessence'' field slow--rolling down its
potential\,\cite{Caldwell} and has adopted many different forms
\cite{PeebRat,Moreq}. The main motivation for the quintessence
models is that they could provide a variable dark energy. Although
there is no compelling reason for a time evolving DE at the moment
-- and it is difficult to get an experimental handle on it
\,\cite{CCvariableTest}--  it is even more difficult to accept (at
least from the theoretical point of view) that the DE can be
described by a static CC term in Einstein's equations with the
small value (\ref{CCvalue}) throughout most of the history of the
universe, namely after the electroweak phase transition took
place.

An alternative  possibility is to think of the DE as a
time-evolving CC. This has been put forward from a purely
phenomenological point of view in many places of the
literature\,\cite{CCphenom,Overduin}. However, a treatment of an
evolving CC term from the more fundamental point of view of the
renormalization group in QFT, with a direct link to Particle
Physics, has been suggested only
recently\,\cite{cosm,JHEPCC1,Babic1,RGTypeIa1}. The possible
experimental consequences of this idea in the light of high-z SNe
Ia data have also been tested in great detail in
\,\cite{RGTypeIa2} (see also the framework and analysis of
\cite{Reuter03a,Bentivegna}). In the renormalization group method
one takes the point of view that $\CC$ should not be constant,
because the quantum effects may shift away the prescribed value,
even if the latter is assumed to be zero\,\cite{cosm}. In this
approach the equation of state for the DE is the one for the
``true'' cosmological term $p_\Lambda = - \rho_\Lambda$. Hence it
is fundamentally different from all kinds of quintessence-like
proposals, where the CC is assumed to vanish precisely and the
role of the DE is played by some new entity with a nontrivial
equation of state. In the RG approach the CC becomes
time-evolving, or redshift dependent: $\CC=\CC (z)$. Although we
do not have a quantum theory of gravity where the running of the
gravitational and cosmological constants could ultimately be
substantiated, a semiclassical description within the well
established formalism of QFT in curved space-time (see e.g.
\cite{birdav,book}) should be a good starting point. From the RG
point of view, $\CC$ becomes a scaling parameter $\CC (\mu)$
whose value should be sensitive to the entire energy history of
the universe -- in a manner not essentially different to, say,
the electromagnetic coupling constant in QED or the gauge
coupling constant in QCD.

While the RG method to tackle the CC problem has been used earlier
\cite{Nelson,Polyakov,lam},  the decoupling effects of the massive
particles at low energies may change significantly the structure
of the renormalization group equations (RGE) with important
phenomenological consequences. This feature has been pointed out
recently by several authors from various interesting points of
view \cite{Babic1,JHEPCC1}. However, it is not easy to achieve a
RG cosmological model where $\CC$ runs smoothly without fine
tuning at the present epoch. In Ref.\,\,\cite{RGTypeIa1,IRGA03} a
successful attempt in this direction has been made, which is based
on possible quantum effects near a high mass scale $M_X$, not far
away from the Planck scale $M_P$, that are transferred (via soft
decoupling) to the low energy physics. At the same time, the
approximate coincidence of the observed CC and the matter
density, $\ \OL^0\sim \OM^0\,$ (i.e. the ``new'' CC problem, or
``time coincidence problem'' \cite{weinRMP,CCRev}) can be
alleviated in this framework\,\cite{IRGA03}. Thus the RG
cosmologies constitute a healthy alternative to the \textit{ad
hoc} quintessence models and are worth exploring\footnote{See
\cite{AHEP03} for a summarized presentation of this RG framework
and an exposition of the basic results. Recently some
applications of the RG method to inhomogeneous cosmologies have
also been proposed\,\cite{Ibanez}.}. In the present paper we
generalize the RG approach developed in
\cite{JHEPCC1,Babic1,RGTypeIa1,IRGA03} by including the
possibility to have a running Newton's constant as
well\,\footnote{Phenomenological models with variable $G$ and
$\CC$, without reference to the RG, have been studied in the
literature since long ago\,\cite{CCphenom, Overduin}. See also the
more recent papers\,\cite{Stefancic}. }. A fundamental question
that we address here is the possibility to maintain the
phenomenologically successful quadratic evolution law
$\delta\CC\sim H^2$ (emerging from an RGE of the form $d\CC/d\ln
H\propto H^2$) as proposed in\,\cite{JHEPCC1,RGTypeIa1,IRGA03},
also in a $G$-running framework. This law is based on the
assumption that the physical RG scale is defined by $\mu=H$, the
Hubble parameter. We will see that this Ansatz is consistent with
the systematic RG scale-setting procedure proposed recently
in\,\cite{Babic2}. In addition, the $\delta\CC\sim H^2$ law is the
only one among those that are consistent with this procedure which
could have measurable phenomenological consequences. Therefore, it
is of foremost importance to explore its phenomenological
viability, in spite of the insurmountable technical difficulties
encountered at present for an explicit calculation in QFT on a
curved background\,\cite{apco}. In this sense it is remarkable
that the quadratic law $\delta\CC\sim H^2$ can also be motivated
from very general arguments in QFT\,\cite{PadmanabhanH2} which may
go beyond the RG. No less appealing is that this quadratic law can
further be claimed, at least heuristically, on the grounds of the
so-called Holographic Principle (HP) applied to effective field
theories\,\cite{Horvat}. In fact, by linking IR and UV scales
through entropy bounds, the HP may provide essential ingredients
to relate the microscopic world of Particle Physics and the large
scale structure of the Universe\,\footnote{See e.g. \cite{Bousso}
for a general review of the HP and its applications.}.  A main
result of this paper is that the adoption of that quadratic law
will lead us to an alternative cosmological RG framework to that
presented in Ref.\,\cite{RGTypeIa1,IRGA03}, keeping though all the
nice phenomenological features of the latter, however with the
distinctive property of having a (logarithmically) running
gravitational constant. This last property endows the new
cosmological model with potentially far-reaching phenomenological
implications both at the cosmological and astrophysical level
which will be explored in some detail in this paper.

The structure of the paper is as follows. In the next section we
introduce a few general features of the cosmological models based
on variable $G$ and $\CC$. In Section \ref{sect:RGmodel1} we
provide some general discussion of the RG cosmological models. The
consistency of the RG  scale setting in cosmology is discussed in
Section \ref{sect:RGmodel3}. In Section \ref{sect:RGmodel2} we
focus on the cosmological model presented here, characterized by a
logarithmic running of $G$ and a quadratic running of $\CC$. The
numerical analysis of this model is presented in Section
\ref{sect:numanalysis}, where we also show the predicted
deviations from the standard FLRW expectations. In Section
\ref{sect:Kepler} we narrow down the applicability of this RG
cosmological model to the astrophysical domain by considering the
possibility to attribute the origin of the flat rotation curves of
the galaxies to a RG quantum correction to Kepler's third law of
celestial mechanics. We close with some final discussion in the
last section, where we draw our conclusions.

\section{Variable $G$ and $\CC$ cosmologies}
\label{sect:VCT}

The cosmological constant contribution to the curvature of
space-time is represented by the $\lambda\,g_{\mu\nu}$ term on
the \textit{l.h.s.} of Einstein's equations.  The latter can be
absorbed on the \textit{r.h.s.} of these equations
\begin{equation}
R_{\mu \nu }-\frac{1}{2}g_{\mu \nu }R=8\pi G\
\tilde{T}_{\mu\nu}\,, \label{EE}
\end{equation}
where the modified $\tilde{T}_{\mu\nu}$ is given by
$\tilde{T}_{\mu\nu}\equiv T_{\mu\nu}+g_{\mu\nu}\,\CC $. Here
$T_{\mu\nu}$ is the ordinary energy-momentum tensor associated to
isotropic matter and radiation, and the new CC term
$\CC=\lambda/8\pi\,G\equiv\rL$ represents the vacuum energy
density. Modeling the expanding universe as a perfect fluid with
velocity $4$-vector field $U^{\mu}$, we have
\begin{equation}
T_{\mu\nu}=-p\,g_{\mu\nu}+(\rho+p)U_{\mu}U_{\nu}\,,
\label{Tmunuideal}
\end{equation}
where $p$ is the proper isotropic pressure and $\rho$ is the
proper energy density of matter-radiation.  Clearly the modified
$\tilde{T}_{\mu\nu}$ defined above takes the same form as
(\ref{Tmunuideal}) with $\rho\rightarrow \tilde{\rho}=\rho+\CC$
and $\ p\rightarrow \tilde{p}=p-\CC$. With this generalized
energy-momentum tensor, and in the FLRW metric ($k=0$ for flat,
$k=\pm 1$ for spatially curved, universes)
\begin{equation}\label{FLRWm}
  ds^2=dt^2-a^2(t)\left(\frac{dr^2}{1-k\,r^2}
+r^2\,d\theta^2+r^2\,\sin^2\theta\,d\phi^2\right)\,,
\end{equation}
the gravitational field equations boil down to Friedmann's
equation
\begin{equation}
H^{2}\equiv \left( \frac{\dot{a}}{a}\right) ^{2}=\frac{8\pi\,G }{3}%
\tilde{\rho} -\frac{k}{a^{2}}=\frac{8\pi\,G }{3}%
\left( \rho +\Lambda\right) -\frac{k}{a^{2}}\,,  \label{FL1}
\end{equation}
and the dynamical field equation for the scale factor:
\begin{equation}
\ddot{a}=-\frac{4\pi}{3}G\,(\tilde{\rho}+3\,\tilde{p})\,a=-\frac
{4\pi}{3}G\,(\rho+3\,p-2\,\CC)\,a\,. \label{newforce3}
\end{equation}
The measured CC is the parameter $\Lambda$ entering equations
(\ref{FL1}) and (\ref{newforce3}), which are used to fit the
cosmological data\,\cite{R98P99,R04P03}.  Let us next contemplate
the possibility that $G=G(t)$ and $\CC=\CC(t)$ can be both
functions of the cosmic time. This is allowed by the Cosmological
Principle embodied in the FLRW metric (\ref{FLRWm}). Then the
Bianchi identities imply that
$\bigtriangledown^{\mu}\,\left({G\,\tilde{T}}_{\mu\nu}\right)=0$,
and with the help of the FLRW metric a straightforward
calculation yields
\begin{equation}\label{BianchiGeneral}
\frac{d}{dt}\,\left[G(\CC+\rho)\right]+3\,G\,H\,(\rho+p)=0\,.
\end{equation}
This equation just reflects the purely geometric Bianchi identity
satisfied by the tensor on the \textit{l.h.s.} of Einstein's
equations (\ref{EE}) in terms of the physical quantities (energy
densities, pressure and gravitational coupling) on its
\textit{r.h.s.} It is easy to check that (\ref{BianchiGeneral})
constitutes a first integral of the dynamical system (\ref{FL1})
and (\ref{newforce3}).  However, this general constraint on the
physical quantities is too complicated in practice, and it is
difficult to understand it on physical grounds because it does not
generally lead to a local energy-momentum conservation law. Let us
consider some simpler cases. The simplest possible case is when
both $G$ and $\CC$ are constant. When $G$ is constant, the
identity above implies that $\CC$ is also a constant, if and only
if the ordinary energy-momentum tensor is individually conserved
($\bigtriangledown^{\mu}\,{T}_{\mu\nu}=0$), i.e.
\begin{equation}\label{NoBronstein}
\dot{\rho}+3\,H\,(\rho+p)=0\,.
\end{equation}
 However, a first non-trivial situation appears when
$\CC=\CC(t)$ but $G$ is still constant. Then a first integral of
the system (\ref{FL1}) and (\ref{newforce3}) is given by
\begin{equation}\label{Bronstein}
\dot{\CC}+\dot{\rho}+3\,H\,(\rho+p)=0\,.
\end{equation}
This result is simpler than (\ref{BianchiGeneral}), and has the
advantage that it can still be interpreted as a physical
conservation law. The conserved (total) energy density here is
the sum of the matter-radiation energy density, $\rho=\rM+\rR$,
plus the dark energy density embodied in $\rL\equiv\CC$. Of
course $\rM$ includes the contribution from the purported cold
dark matter (CDM), which should be dominant. Alternatively, using
the metric (\ref{FLRWm}) one may derive (\ref{Bronstein})
directly from $\bigtriangledown^{\mu}\,{\tilde{T}}_{\mu\nu}=0$, as
this is obviously a particular case of (\ref{BianchiGeneral}).
This possibility shows that a time-variable CC cosmology may exist
such that transfer of energy occurs from matter-radiation into
vacuum energy, and vice versa. From the point of view of general
covariance there is no problem because the latter is tied to the
conservation of the full energy-momentum tensor, which in this
case is $\tilde{T}_{\mu\nu}$ rather than ${T}_{\mu\nu}$. We
remark that while the decay of the vacuum energy into ordinary
matter and radiation could be problematic for the observed CMB,
the decay into CDM could be fine\,\cite{PeebRat,OpherPel}.  The
detailed analysis of a time-variable CC cosmology in the light of
high-z SNe Ia data carried out in Ref.\,\cite{RGTypeIa2} was
precisely based on this framework.

In this paper we make a step forward by entertaining the
possibility that $G$ can also be a function of the cosmic time,
$G=G(t)$, together with $\CC=\CC(t)$. Although we have seen that
the most general case should fulfill the general constraint
(\ref{BianchiGeneral}), we still want to keep the usual
matter-radiation conservation law (\ref{NoBronstein}) as this
represents the canonical (i.e. the simplest) cosmological model
with variable $G$ and $\CC$ possessing a physical conservation
law. In this model there is no transfer of energy between
matter-radiation and dark energy. However, the time evolution of
$\CC$ is still possible at the expense of a time varying
gravitational constant. Indeed, it is easy to see that in this
case the following (hereafter called canonical) differential
constraint must be satisfied:
\begin{equation}\label{difCCG}
(\rho+\CC)\,\dot{G}+G\,\dot{\CC}=0\,.
\end{equation}
Again this is a particular case of (\ref{BianchiGeneral}). The
solution of a generic cosmological model of this kind with
variable $G$ and $\CC$ is contained in part in the coupled system
of differential equations (\ref{FL1}), (\ref{NoBronstein}) and
(\ref{difCCG}), together with the equation of state $p=p(\rho)$
for matter and radiation. However, still another equation is
needed to completely solve this cosmological model in terms of
the fundamental set of variables $(H(t),\rho(t),p(t), G(t),
\CC(t))$. Many relations have been proposed in the literature
from the purely phenomenological point of
view\,\cite{CCphenom,Overduin}. In the next section we discuss
the possibility that the fifth missing equation is a relation of
the form $\CC=\CC(H)$. Most important, we will motivate this
relation at a fundamental QFT level, specifically within the
context of the renormalization group in curved space-time, and
then discuss the possible form of the associated RG evolution law
for the gravitational constant.

\section{Renormalization group cosmologies}
\label{sect:RGmodel1}

Following the approach of \cite{JHEPCC1,Babic1,RGTypeIa1,IRGA03}
we describe the scaling evolution of $\CC$  by introducing a
renormalization group equation for the cosmological constant. The
connection between scaling evolution and time evolution is to be
made in a second step. From dimensional analysis, and also from
dynamical features discussed in the aforementioned references,
the RGE  may take in principle the generic form
\begin{equation}\label{RGEG1a}
\frac{d\CC}{d\ln\mu}= \sum_n\,A_n\,\mu^{2n}\,.
\end{equation}
Here $\mu$ is the energy scale associated to the RG running.
Furthermore, in a RG cosmology with a time-evolving $G$ we expect
a RGE for $G$ as well.  This is best formulated in terms of
$G^{-1}$ as follows:
\begin{equation}
\frac{d}{d\ln\mu}\left(\frac{1}{G}\right)=
\sum_n\,B_n\,\mu^{2n}\,. \label{RGEG1b}
\end{equation}
It will prove convenient for further reference to introduce the
integrated form of these equations:
\begin{equation}\label{RGEInt}
\CC(\mu)= \sum_n\,C_n\,\mu^{2n}\,,\ \ \ \ \ \  \frac{1}{G(\mu)}=
\sum_n\,D_n\,\mu^{2n}\,.
\end{equation}
Coefficients $A_{n}, B_{n}, C_{n}$ and $D_{n}$ in these formulas
are obtained from a sum of the contributions of fields of
different masses $M_i$. The \textit{r.h.s.} of (\ref{RGEG1a}) and
(\ref{RGEG1b}) define the $\beta$-functions for $\CC$ and $G^{-1}$
respectively, which are functions of the masses and in general
also of the ratios between the RG scale $\mu$ and the masses. For
$\mu\ll M_i$, the series above are in principle assumed to be
expansions in powers of the small quantities
$\mu/M_i$\,\cite{JHEPCC1,Babic1}. However, the possibility to
extend the analytical properties of some of these expansions will
be discussed later. We assume that $\mu$ is of the order of some
physical energy-momentum scale characteristic of the FLRW
cosmological setting, and can in principle be specified in
different ways.  In the $\CC$-running framework of \cite{JHEPCC1},
it was assumed that $\mu$ is given by the typical energy-momentum
of the cosmological gravitons of the background FLRW metric. This
suggests we can set $\mu\simeq H$ (Hubble parameter), which is of
the order of the square root of the curvature scalar in the FLRW
metric, $\mu\sim R^{1/2}$.

In the cosmological applications, e.g. when considering the
evolution of the present universe, we are going to keep the
Ansatz $\mu\simeq H$. We will claim that it is a sound RG
scale-setting also for a framework based on running $\CC$
\textit{and} $G$. The consistency of this choice will be discussed
in Section \ref{sect:RGmodel3} in the light of the scale-setting
procedure devised in \cite{Babic2}. For the moment we observe
that, within the $\mu\simeq H$ Ansatz, the RGE (\ref{RGEG1a})
above defines a relation of the type $\CC=\CC(H)$, and therefore
provides the fifth equation suggested at the end of the previous
section, needed to solve the cosmological model. At the same time
Eq.\,(\ref{RGEG1b}) defines a relation of the sort $G=G(H)$.
Obviously these two functions of $H$ cannot be independent, for
the Bianchi identity imposes the differential constraint
(\ref{difCCG}) in which $\rho=\rho(t)$ is a known function
obtained after solving (\ref{NoBronstein}) -- once the equation of
state $p=p(\rho)$ for matter and radiation is specified. Put
another way: consistency requires that the two expansions
(\ref{RGEG1a}) and (\ref{RGEG1b}) ought to be interconnected via
the differential constraint (\ref{difCCG}), so that if one of the
expansions is given the other can be retrieved from the Bianchi
identity. For example, in Ref.\,\cite{RGTypeIa1} we suggested the
following specific form for the RGE (\ref{RGEG1a})
\begin{equation}\label{RG01}
\frac{d\CC}{d{\ln \mu}}= \frac{1}{(4\pi)^{2}}\
\sigma\,\mu^{2}M^{2}+...\,,
\end{equation}
with $\mu=H$. Here we have defined the effective mass parameter
\begin{equation}\label{Mdef}
M\equiv\left|\sum_i\,c_i\,M_i^2\right|^{1/2}\,,
\end{equation}
and $\,\sigma=\pm 1$ indicates the sign of the overall
$\bCC$-function, depending on whether the fermions ($\sigma=-1$)
or bosons ($\sigma=+1$) dominate at the highest energies. The
form (\ref{RG01}) for the RGE is based on the leading
participation of the heaviest possible masses $M_i$ in the
expansion. These masses could arise e.g. from some Grand Unified
Theory (GUT) whose energy scale $M_X$ is just a few orders of
magnitude below $M_P$. The form (\ref{RG01}) results from
truncating the series (\ref{RGEG1a}) at $n=1$. The term $n=0$,
corresponding to contributions proportional to $M_i^4$, must be
absent if we want to describe a successful
phenomenology\,\cite{RGTypeIa1,RGTypeIa2}. Actually, from the RG
point of view it is forbidden because in the physics of the
modern Universe we probe a range of energies where $\mu$ is
always below the superheavy masses $M_i$. In addition,
dimensional analysis and dynamical expectations tell us that the
remaining terms are suppressed by higher powers of
$\mu/M_i$\,\,\cite{JHEPCC1,Babic1,RGTypeIa1}. In the absence of
experimental information, the numerical choice of
$\,\sigma\,M^2\,$ is model-dependent. For example, the fermion and
boson contributions in (\ref{RGEG1a}) might cancel due to
supersymmetry (SUSY) and the total $\beta$-function becomes
non-zero at lower energies due to SUSY breaking.  Another option
is to suppose some kind of string transition into QFT at the
Planck scale. Then the heaviest particles would have masses
comparable to the Planck mass $M_P$ and represent the remnants,
e.g., of the massive modes of a superstring.

The obvious motivation for this RG
scenario\,\cite{RGTypeIa1,IRGA03} is that the \textit{r.h.s.} of
(\ref{RG01}) is of the order of the present value of the CC. To
assess this in more detail, let us integrate Eq.\,(\ref{RG01})
straightforwardly. We find an evolution law for $\CC$ which is
quadratic in $\mu$\, and can be expressed, with the help of the
notation (\ref{RGEInt}), as follows:
\begin{equation}\label{CCH}
\CC(\mu)=C_0+C_1\,\mu^2\,,\\
\end{equation}
where
\begin{equation}\label{C0C1}
C_0=\Lambda_0-\frac{3\,\nu}{8\pi}M_P^2\,\mu_0^2\,, \ \ \
C_1=\frac{3\,\nu}{8\pi}\,M_P^2\,.
\end{equation}
Within our Ansatz for the RG scale, $\mu_0$ stands for the present
day value of the Hubble parameter, $H_0$, and the coefficient
$\nu$ is essentially the ratio between the squares of the
effective GUT mass scale $M$ and the Planck scale up to
$\sigma=\pm 1$, depending on whether bosonic or fermionic fields
dominate in the mass spectrum below the Planck scale:
\begin{equation}\label{nu}
\nu=\frac{\sigma}{12\pi}\frac{M^2}{M_P^2}\,.
\end{equation}
This parameter was first defined in
Ref.\,\cite{RGTypeIa1,RGTypeIa2}. Irrespective of the precise
meaning of $\nu$ in the context of QFT in curved space-time, the
latter reference suggested also how to use $\nu$ to parametrize
deviations of the evolution laws from the standard FLRW
cosmology\,\footnote{See e.g. Eq.\,(4.11)-(4.14) of
\cite{RGTypeIa2} and discussions therein. Recently, this purely
phenomenological point of view has been re-taken in \cite{Wang}
where a similar parameter is called $\epsilon$ ($=3\,\nu$).}.
Equation\,(\ref{CCH}) is normalized such that $\CC$ takes its
present day value $\CC_0$ for $\mu=\mu_0$. The highest possible
value of the heavy masses $M_i$ is $M_P$, and therefore we
typically expect that the effective mass (\ref{Mdef}) is of the
same order.  However, due to the multiplicities of the heavy
masses, we cannot exclude a priori that $M>M_P$. To be
conservative we will take
\begin{equation}\label{nu0}
|\nu|=\nu_0\equiv\frac{1}{12\pi}\simeq 0.026
\end{equation}
as the typical value of $\nu$, corresponding to $M=M_P$. A very
nice feature of this RG picture for $\CC$ is that its evolution,
being smooth and sufficiently slow in the present Universe, may
still encapsulate non-negligible effects that could perhaps be
measured\,\footnote{One could not achieve this extremely welcome
pay-off if using RG scales other than $\mu\sim H$ (e.g.
$\mu\sim\rho^{1/4}$) and particle masses of order of those in the
SM of particle physics. One unavoidably ends up with fine tuning
of the parameters. See the detailed discussions in
\,\cite{JHEPCC1,Babic1}.}. In fact, the predicted rate of change
of $\CC$ (in other words, the value of its $\beta$-function) at
$\mu = H_0$  is of order of $\CC_0$ itself:
\begin{equation}
\beta_{\Lambda}^0\sim
C_1\,H^2\left|_{t=t_0}\right.=\frac{3\,\nu}{8\pi}\,M_P^2\,H_0^2\lesssim
10^{-47}\,GeV^4\,. \label{betaPlanck}
\end{equation}
In contrast, at the earliest times when $H\simeq M_P$, the same
RGE predicts a CC of order $\CC\sim M_P^4$, which again is a
natural expectation.  If we define the energy scale
$E_{\CC}(t)\equiv\CC(t)^{1/4}$ associated to the vacuum energy
density $\rL\equiv\Lambda$ at any given moment of the history of
the Universe, the RGE above may explain\,\cite{RGTypeIa1,IRGA03}
why the millielectronvolt energy associated to the present value
$\CC_0$ happens to be the geometrical mean of the two most
extreme energy scales in our Universe, $H_0\sim 10^{-42}\,GeV$
(the value of $\mu$ at present) and $M_P\sim 10^{19}\,GeV$:
\begin{equation}\label{millielectronvolt2}
E_{\CC}(t=t_0)\equiv\CC_0^{1/4}\simeq\sqrt{M_P\,H_0}={\cal O}
\left(10^{-3}\right)\,eV\,.
\end{equation}
Amusingly this is also the order of magnitude of another,
independent, cosmic energy scale: the one associated to matter
density, $E_M(t)\equiv\rM^{1/4}(t)$. Assuming $k=0$ and $\CC=0$
in Friedmann's equation (\ref{FL1}), the energy scale associated
to the matter density at present is seen to be $E_M(t_0)\sim
\sqrt{M_P\,H_0}$. Remarkably the dark energy scale
(\ref{millielectronvolt2}) associated to our RG cosmological
framework coincides with the matter density scale defined by
Friedmann's equation at $t=t_0$, and therefore it helps to
explain the so-called ``coincidence problem''\,\cite{CCRev}, one
of the two fundamental conundrums behind the cosmological
constant problem.

A value of $\nu$ of order of $10^{-2}-10^{-3}$ suggests that
there might be some physics just 2 orders of magnitude below the
Planck scale, i.e. around $10^{17}\,GeV$. This would be indeed
the mass scale of the heavy  masses $M_i$ if we assume some
multiplicity of order $100$ in the number of the heaviest degrees
of freedom in a typical GUT. To see this, rewrite (\ref{nu}) with
$\nu=10^{-3}$ as
\begin{equation}\label{MX}
\sum_i c_i\,M_i^2=12\,\pi\,M_P^2\,\nu= 12\,\pi\,(1.2\times
10^{19}\,GeV)^2\,10^{-3}\simeq 5.4\times 10^{36}\,GeV^2\,.
\end{equation}
Then if we assume $M_i=M_X$, $c_i\sim 1$ for all $i$, it follows
that
\begin{equation}\label{MX2}
M_X\sim 10^{17}\,GeV\  \ \ {\rm  for}\ \ \ \sum_i\,c_i\sim 100\,.
\end{equation}
Therefore, there is the possibility that the origin of this RG
cosmology could bare some relation to physics near the SUSY-GUT
scale.

From the previous considerations we realize that the models based
on the RG evolution of $\CC$ could be serious competitors to e.g.
quintessence models (and generalizations thereof) to account for
some of the fundamental issues behind the multifarious
cosmological constant problem\,\cite{weinRMP,CCRev}, and moreover
they are also testable in the next generation of  high-z SNe Ia
experiments (such as SNAP\cite{SNAP}) -- as has been demonstrated
in great detail in \,\cite{RGTypeIa2}. The RG cosmological model
analyzed in the last reference, although it constitutes a running
$\CC$ model, is a fixed-$G$ model.  In contrast, the RG model
under consideration aims at preserving the virtues of the
quadratic law (\ref{CCH}) for the $\CC$ evolution (i.e. the rich
phenomenology potential unraveled in \,\cite{RGTypeIa2}) while
allowing for a simultaneous evolution of the gravitational
constant, $G$. This may add distinctive features and new
phenomenological tests to be performed. Obviously it is important
to analyze this possibility in order to distinguish between these
two RG cosmological models and also to fully display the power of
the RG approach versus the other dark energy models\,\cite{CCRev}.

\section{On the RG scale-setting procedure in cosmology}
\label{sect:RGmodel3}

In this section we consider the consistency of the RG scale choice
in certain problems related to the cosmological framework.
Specifically, we discuss the consistency of our choice $\mu\simeq
H$ for the application to the large scale cosmological problems
along the lines of the procedure recently proposed
in\,\cite{Babic2}. In the next two sections we shall concentrate
on the construction of a cosmological model based on the RG
approach, and only in the last part of our work we will try to
extend these methods to the local (galactic) domain. Let us
commence with the scale setting at the cosmological level.
Starting from the canonical constraint (\ref{difCCG}) we trade
the cosmic time for a generic scale parameter $\mu$ (without
specifying its relation to physics at this point), and rewrite it
in the following way\,\cite{Babic2}:
\begin{equation}\label{difCCG2}
\left[(\rho+\CC
(\mu))\,\frac{dG(\mu)}{d\mu}+G(\mu)\,\frac{d\CC(\mu)}{d\mu}\right]\,\frac{d\mu}{dt}=0\,.
\end{equation}
Obviously, the matter-radiation energy density $\rho$ is the only
parameter which can not be defined as a function of $\mu$ directly
because it is unrelated to the RG in the first place. However,
since $G$ and $\Lambda$ are primarily functions of $\mu$ in the RG
framework, the differential constraint (\ref{difCCG2}) can be used
to eventually define $\rho$ as a function of $\mu$ . If
$d\mu/dt\neq 0$, it is easy to check that we can isolate $\rho$ as
follows:
\begin{equation}\label{rhoimplicit}
\rho=-\Lambda(\mu)+\frac{1}{G(\mu)}\,\frac{d\CC(\mu)}
{d\mu}\left(\frac{d}{d\mu}\,\frac{1}{G(\mu)}\right)^{-1}\,.
\end{equation}
This scale-setting algorithm defines a relation  $\rho=f(\mu)$
between the cosmological data, represented by $\rho$, and the
scaling functions of $\mu$. Under appropriate conditions, this
relation can be inverted to find $\mu$ as a function of the
external data: $\mu=f^{-1}(\rho)$. Since this function has been
determined by the Bianchi identity, it should define a valid set
of RG scales for cosmology. In other words, the specific scale
$\mu$ to be used in cosmology should be chosen among the class of
(potentially multivalued) functions obtained by this procedure
under the condition that the scale so obtained is
positive-definite and is a smooth function of the cosmological
parameters. Further (physical) criteria may be needed to
ultimately select one particular function within this class.

Assuming that this is a valid method to define the RG scale, we
are now in position to develop a scenario with simultaneous
running of the parameters $G$ and $\CC$ in which the choice
$\mu\simeq H$ is compatible with the aforementioned scale-setting
procedure. To realize this scenario we stick to our RGE for $\CC$
as given in (\ref{RG01}) -- hence to the quadratic running law
(\ref{CCH}) --  and we seek for a suitable RGE for $G^{-1}$. It
cannot be an independent equation once the RGE for $\CC$ has been
established. Therefore, this restriction provides a method to
hint at possible forms for the RGE (\ref{RGEG1b}) compatible with
the canonical constraint (\ref{difCCG}). For example we could
start with an expansion akin to that in Eq.\,(3.1) of
\,\cite{RGTypeIa2} for $\CC$:
\begin{equation}\label{RGEG}
(4\pi)^2\,\frac{d}{d\ln\mu^2}\left(\frac{1}{G}\right)=
\sum_i\,a'_i\,M_i^2+\sum_i\,b'_i\,M_i^2\left(\frac{\mu}{M_i}\right)^2+
\sum_i\,c'_i\,M_i^2\left(\frac{\mu}{M_i}\right)^4+...
\end{equation}
Since the cosmological RG scale $\mu$ is assumed to be very small
for the present universe, and the series is supposed to be
convergent, we expect that the first term in this expansion is
the crucial one and the remaining terms are inessential. Indeed,
selecting the first term we are led to a RGE of the type
\begin{equation}\label{RGEG22}
\frac{d}{d\ln \mu^2}\left(\frac{1}{G}\right)= \,\nu\,M_P^2+...\,.
\end{equation}
Consider now the generic expansions (\ref{RGEG1a})-(\ref{RGEInt})
for $\CC$ and $G^{-1}$. In particular notice the following
relations among the coefficients for $n=0$:
\begin{eqnarray}\label{RGEG2}
&&C_0=\CC_0+A_0\,\ln\frac{\mu}{\mu_0}-\frac12\,A_1\,\mu_0^2-\frac14\,A_2\,\mu_0^4+...
\nonumber\\
&&D_0=M_P^2+B_0\,\ln\frac{\mu}{\mu_0}-\frac12\,B_1\,\mu_0^2-\frac14\,B_2\,\mu_0^4+...\,,
\end{eqnarray}
where $\mu_0$ is the normalization point corresponding to the
$t=t_0$ moment of time. The coefficients (\ref{RGEG2}) play a role
in computing explicitly the \textit{r.h.s.} of
(\ref{rhoimplicit}). In particular notice that $A_0=0$, to be
consistent with (\ref{C0C1})(cf. Ref.\,\cite{RGTypeIa1}). Also
important is that the scale derivative of $D_0$ is non-vanishing,
\begin{equation}\label{D0P}
D_0'\equiv\frac{dD_0}{d\mu}=\frac{B_0}{\mu}
=\frac{\sigma}{6\pi}\,\frac{M^2}{\mu}\,.
\end{equation}
Here we have used $B_0=2\,\nu\,M_P^2=\sigma\,M^2/6\,\pi$ as seen
from (\ref{RGEG22}) and (\ref{nu}). This result shows that
$G^{-1}$ is actually non-analytic at $\mu=0$, a fact to which we
shall return later. With these relations in mind, a
straightforward calculation of the \textit{r.h.s.} of
Eq\,(\ref{rhoimplicit}), using the expansions (\ref{RGEInt}),
leads to
\begin{equation}\label{rhosolved}
\rho\simeq -C_0+C_1\left(\frac{2\,D_0}{B_0}-1\right)\,\mu^2+...
\end{equation}
where we have neglected the higher order terms in $\mu$. Next we
substitute the coefficients (\ref{C0C1}) in (\ref{rhosolved}) and
use the fact that $D_0\simeq M_P^2$. Irrespective of the sign
$\sigma=\pm 1$, the resulting equation determining $\mu$ is
amazingly simple:
\begin{equation}\label{muH}
\mu^2\simeq
\frac{\rho+C_0}{C_1\left(\frac{2\,M_P^2}{B_0}-1\right)}
\simeq\frac{8\pi}{3}\,\frac{\rho+\CC_0}{M_P^2}=H^2\,,
\end{equation}
where the last equality makes use of Friedmann's equation
(\ref{FL1}) in the flat case. Our starting Ansatz $\mu=H$ is,
therefore,  explicitly borne out within the criterion defined by
Eq.\,(\ref{rhoimplicit}) \ (\textit{q.e.d.}).

The previous proof relies on the expansions
(\ref{RGEG1a})-(\ref{RGEInt}) advocated in the
literature\,\cite{JHEPCC1,Babic1,RGTypeIa1,Babic2}.
Notwithstanding, this does not exhaust the list of possibilities
made available to us by the Bianchi identity in the canonical
form (\ref{difCCG}). It is still possible to extract the same
phenomenology if we make allowance for an extended RGE for $G$
beyond an expansion in (integer) powers of the small parameters
$\mu/M_i$. Then other (non-analytic) contributions may arise.  We
shall discuss this possibility along the second part of the next
section.


\section{A cosmological model with logarithmic running of G and
 quadratic running of $\CC$}
\label{sect:RGmodel2}

In the previous section we have established a scenario where the
choice $\mu\simeq H$ is a consistent scale setting for
cosmological considerations. Let us use this RG framework and
apply it to a spatially flat Universe, i.e. $k=0$ in
Eq.\,(\ref{FL1}). This situation is not only the simplest one,
but it corresponds (as already mentioned in the introduction) to
the preferred scenario in the light of the present data from
distant supernovae and the CMB. Moreover, the $k=0$ Universe
already captures all the main traits of the new RG cosmology with
variable $G$ without introducing additional complications related
to curvature. By collecting Friedmann's equation, the quadratic
RG evolution law for $\CC$, and the (canonical) differential
constraint imposed on the functions $G(t)$ and $\CC(t)$ by the
Bianchi identity, we can set up the system:
\begin{eqnarray}\label{System1}
&&\rho+\CC=\frac{3H^2}{8\,\pi\,G} \nonumber\,,\\
&& \CC=C_0+C_1\,H^2\,,\\
&&(\rho+\CC)\,d{G}+G\,d{\CC}=0\nonumber\,.
\end{eqnarray}
This system can be immediately solved for the function
$G=G(H;\nu)$ depending of the parameter $\nu$, with the result
\begin{equation}\label{GH}
G(H;\nu)=\frac{G_0}{1+\nu\,\ln\left(H^2/H_0^2\right)}\,,
\end{equation}
where $G(H_0)=G_0\equiv 1/M_P^2$. In this way we have obtained
explicitly the corresponding running law for $G$ as a function of
the RG scale $\mu=H$. From this law we learn that for $\nu>0$ the
gravitational constant $G$ decreases logarithmically with $H$,
hence very slowly, which is an extremely welcome feature in view
of the experimental restrictions on $G$ variation (see later on).
Such a decreasing behavior with the cosmological energy scale
$\mu=H$ is equivalent to say that the law that we have obtained
nicely corresponds to an asymptotically free regime for $\nu>0$,
much in the same way as the strong coupling constant in QCD. In
addition, we discover that the parameter $\nu$ defined in
(\ref{nu}) plays essentially the role of the (dimensionless)
$\beta$-function for $G$. Finally, we note that the function
(\ref{GH}) is non-analytic, which was expected from a remark made
in Section \ref{sect:RGmodel3}.

Consistency with the Bianchi identity requires that
Eq.\,(\ref{GH}) must be the solution of the RGE (\ref{RGEG1b})
for $G^{-1}$. To check this, we note that in leading order the
last equation boils down to Eq.\,(\ref{RGEG22}).  Integrating
this equation under the boundary condition $G(\mu=\mu_0)=G_0$ we
immediately recover Eq.\,(\ref{GH}), as desired. Had we ignored
the first term in (\ref{RGEG}) and assumed that the first leading
contribution comes from the second term (i.e. the one
proportional to $\mu^2=H^2$) we would have obtained
\begin{equation}\label{GRun2}
G(H)=\frac{G_0}{1+\zeta\,G_0\,(H^2-H^2_0)}\,,
\end{equation}
instead of (\ref{GH}), where $\zeta$ is a dimensionless
coefficient. The resulting scenario is not very appealing as it
does not lead to any phenomenological consequence to speak of. In
fact, if we substitute (\ref{GRun2}) in (\ref{difCCG}) and
integrate the resulting equation, the following evolution law for
the CC ensues:
\begin{equation}\label{CCRun2}
\CC (H)=\CC_0 +\frac{3\,\zeta}{16\pi}\,(H^4-H_0^4)\,.
\end{equation}
Although this quartic law is -- as the quadratic one (\ref{CCH})--
smooth and mathematically possible, it leads to absolutely no
phenomenology due to the smallness of the Hubble parameter at the
present time: $H_0\sim 10^{-42}\,GeV$. Similarly, the square
dependence $G_0\,H^2=H^2/M_P^2$ in the running law for $G$ given
by (\ref{GRun2}) is extremely weak. Therefore, it is clear that
only if the first term of the \textit{r.h.s.} of (\ref{RGEG}) is
non-vanishing it is possible to meet the logarithmic law
(\ref{GH}) for $G$ and the quadratic law (\ref{CCH}) for $\CC$.
This combination of RG laws is the only one compatible with the
canonical constraint (\ref{difCCG}) that may lead to a
cosmological model with a rich phenomenology comparable to the
model presented in \cite{RGTypeIa1,RGTypeIa2}.

A remark is now in order. As the reader could readily notice, by
admitting that the first term of the \textit{r.h.s.} of
(\ref{RGEG}) is non-vanishing, we are departing from the
decoupling theorem\,\cite{AC} and assume the non-standard form of
decoupling for the parameter $1/G$. In order to justify this
departure from the canonical form of decoupling, let us remember
that the attempt to derive the gravitational version of the
decoupling theorem\,\cite{apco} was successful only in that
sector of the gravitational action which is available in the
linearized gravity framework. In the cases of $\CC$ and $1/G$ the
explicit derivation of the decoupling law requires a new
technique of calculation which should take the non-trivial and
dynamical nature of the metric background into account. In the
actual phenomenological approach it is primarily the Bianchi
identity (rather than the decoupling theorem) that decides about
the evolution of $G$ once the evolution of $\CC$ has been
ordered. In this sense, the final structure for the RGE of $G$
can be of the form (\ref{RG01}) even if this does not fulfill the
standard expectations from the decoupling theorem. This uneven
status between the two parameters can be understood on physical
grounds, in the sense that the ultimate origin of $G$ and $\CC$
and their role in defining the background geometry may be
essentially different.  $\CC$ can be viewed as a small parameter
in QFT: this can be technically implemented by allowing for a
fine-tuning of the renormalization condition for the vacuum
counterpart of the cosmological term -- see the Introduction. At
the same time the situation with $1/G$ is quite different. For
example, it could be that the Planck mass is not only some
threshold scale for the quantum gravity effects but, after all, a
real mass. One can imagine that the non-perturbative phenomena of
quantum gravity are involved to generate this scale and then the
standard decoupling law (which is essentially a perturbative
result) would not apply to the parameter $1/G$.

We shall finish this section by presenting additional arguments
concerning our choice for the identification of the scale
parameter $\mu$ with the Hubble parameter $H$ at low energies.
These arguments do not look superfluous in view of the distinct
choices for $\mu$ which one can meet in the existing
literature\,\cite{cosm,JHEPCC1,Babic1,RGTypeIa1,Reuter03a,Babic2}.
If one allows for some generalization of the expansions
(\ref{RGEInt}), it may have an impact on the scale setting. It is
thus advisable not to assume a priori the result $\mu=H$ derived
in (\ref{muH}), although we want to show that under certain
conditions we can also retrieve a similar scale setting. On
physical grounds we cannot exclude a priori the presence of
non-perturbative and/or non-local effects in the far
IR\,\cite{Woodard}. Whereas at higher energies the behavior of
$G(\mu)^{-1}$ is influenced by the higher powers of $\mu$ in
(\ref{RGEInt}), the running in the present day Universe is
controlled by the first coefficient $B_0$ in (\ref{RGEG1b}),
which in turn leads to a non-analytical behavior of $D_0=D_0
(\mu)$ as a function of $\mu$ in the series (\ref{RGEInt}). In
other words, at the present IR energies the expansion $G^{-1}$ in
powers of $\mu$ may not be a pure Taylor expansion.  Baring in
mind these considerations, let us generalize the picture as
follows. Assume that in the far IR the expansions (\ref{RGEInt})
take contributions of the form
\begin{equation}\label{CCGgeneric}
\CC(\mu)=a_0+a_1 \left(\frac{\mu}{\mu_0}\right)^{\alpha}\,,\ \ \
\frac{1}{G(\mu)}=b_0+b_1\left(\frac{\mu}{\mu_0}\right)^{\beta}\,,
\end{equation}
where we assume arbitrary coefficients $a_0,b_0,a_1,b_1$ and
arbitrary exponents $\alpha,\beta$, which can be smaller than
one. From these general expressions we may compute the
corresponding RG scale following from (\ref{rhoimplicit}). At the
end of the calculation we encounter the result:
\begin{equation}\label{muGeneral}
\rho+a_0=\frac{a_1}{\beta}\left[(\alpha-\beta)\left(\frac{\mu}{\mu_0}\right)^{\alpha}
+\alpha\,\,\frac{b_0}{b_1}\,\left(\frac{\mu}{\mu_0}\right)^{\alpha-\beta}
\right]\,.
\end{equation}
This example already shows that even in relatively simple
situations it may be difficult to find the explicit relation
$\mu=f^{-1}(\rho)$ mentioned in Section \ref{sect:RGmodel3}. It
also shows that the scale setting $\mu\simeq H$ will not always be
possible. And, most important, it also helps us to find situations
when it is actually possible. Indeed, let us consider a generic
case where the RGE for $G^{-1}$ takes terms of the form
\begin{equation}\label{dG1}
 \frac{d}{d\ln \mu^2}\ \left(\frac{1}{G}\right)=
\nu\,M_P^2\,\left(\frac{\mu^2}{M^2}\right)^{\gamma}\,.
\end{equation}
For any $\gamma>0$ (integer or not) this expression does respect
the decoupling theorem, in contrast to (\ref{RGEG}).  By
integrating (\ref{dG1}) we are led to an expression for
$G(\mu)^{-1}$ as in (\ref{CCGgeneric}), with
\begin{equation}\label{Gintegral2}
\beta=2\,\gamma\,,\ \ \
b_0=\frac{1}{G_0}\left\{1-\frac{\nu}{\gamma}\left(\frac{\mu_0^2}{M^2}\right)^{\gamma}\right\}\,,
\ \
b_1=\frac{1}{G_0}\,\frac{\nu}{\gamma}\left(\frac{\mu_0^2}{M^2}\right)^{\gamma}\,.
\end{equation}
Together with (\ref{dG1}) let us assume that the $\CC$-scaling law
is given by (\ref{CCGgeneric}) with $\alpha=2$. Then in the limit
$\gamma\ll 1$ (hence $\beta\ll 1$) we may solve approximately
Eq.\,(\ref{muGeneral}) for the RG scale, and we find
\begin{equation}\label{muexample1}
{\mu}^{2}\simeq {\mu_0}^{2}\,\frac{\beta\,b_1}{2\,a_1}\
\frac{\rho+a_0}{b_0+b_1}\simeq\frac{8\,\pi}{3\,M_P^2}\,(\rho+\CC_0)=
H^2\,.
\end{equation}
In deriving this formula we have used: i) the explicit form of the
coefficients in Eq.\, (\ref{Gintegral2}); ii) $a_0=C_0\simeq\CC_0$
and $a_1=\mu_0^2\,C_1$, which follow after comparing
(\ref{CCGgeneric}) with (\ref{CCH})-(\ref{C0C1}); iii)
$\beta\,b_1\simeq 2\nu\,M_P^2$, which holds in the limit
$\gamma\ll 1$. We have thus met again the scale $\mu\simeq H$. We
see that it can be obtained within the more generic class of RGE's
for $G$ of the form (\ref{dG1}) in the limit of small $\gamma$
(corresponding to an effective logarithmic behavior), and when
the cosmological term evolves quadratically with $\mu$, i.e.
$\alpha=2$ in (\ref{CCGgeneric}). This more general analysis may
suggest the following interpretation. Recall that the
cosmological domain has a natural finite size, which is defined
by the largest cosmological distance where we can perform
physical measurements and where all quantum fluctuations remain
confined \,\cite{PadmanabhanH2}: to wit, the horizon $d_h\sim
1/H$.  The energy scale associated to the horizon is precisely
our running scale $\mu=H$. We do not consider graviton momenta
below this scale. The underlying effective action may develop an
IR tail of non-analytic contributions related to non-perturbative
and/or non-local effects that cannot be described by a pure
Taylor expansion. We have modeled this possibility with a RGE for
$G$ of the form (\ref{dG1}) with small $\gamma$. In general we
expect non-analytic effects also for the local systems at, say,
the galactic scale, which are always finite-sized by definition
--see more details on this matter in Section \ref{sect:Kepler}.

In summary, in this section we have shown that the quadratic law
(\ref{CCH}) for $\CC$ and the logarithmic law (\ref{GH}) for $G$
are a pair of RG scaling laws that can be consistently formulated
in terms of the scale $\mu=H$. This is tantamount to say that
these running laws $G=G(\mu)$ and $\CC=\CC(\mu)$, when formulated
in terms of $\mu=H$, can be consistently transformed into
time-evolving functions $G=G(t)$ and $\CC=\CC(t)$ that satisfy the
original (time-dependent) canonical differential constraint
(\ref{difCCG}) reflecting the Bianchi identity of the Einstein
tensor.  Furthermore we have seen that the RG scale for these
laws can be derived from the recipe (\ref{rhoimplicit}) either by
assuming the expansion (\ref{RGEG1b}) with $B_0\neq 0$ as the
leading coefficient, or from the exponential form (\ref{dG1}) in
the limit of small $\gamma$; in both cases it amounts to a
logarithmic behavior of $G=G(\mu)$. As we have advanced in Section
\ref{sect:RGmodel1}, a quadratic law for $\CC$ is very important
for potential phenomenological implications. In the following
section we explore some of these implications for cosmology,
together with the new feature introduced by the logarithmically
varying gravitational constant.\\


\section{Numerical analysis of the RG cosmological model}
\label{sect:numanalysis}

The RG cosmological model under consideration is based on the
system of equations (\ref{System1}) and the ordinary conservation
law (\ref{NoBronstein}) for matter and radiation. While the
solution has already been exhibited in the form $G=G(H;\nu)$,
$\CC=\CC(H;\nu)$ in Section \ref{sect:RGmodel2}, we want to
present also the solution in terms of the redshift variable,
$z=(a_0-a)/a$, following a procedure similar to
\cite{RGTypeIa1,RGTypeIa2}. This is specially useful for
phenomenological applications, in particular to assess the
possibility that this model can be tested in future cosmological
experiments such as SNAP\,\cite{SNAP}. Therefore, we look for the
one-parameter family of functions of the variable $z$ at a given
value of $\nu$:
\begin{equation}\label{functionsz}
G=G(z;\nu)\,,\ \ \ \ \ \ \ \ \CC=\CC(z;\nu)\,.
\end{equation}
We are going to analyze first in detail the solution
(\ref{functionsz}) for the present (matter-dominated) universe,
where $\rho=\rM$, $\rR\simeq 0$, and then make some comments on
the implications for the nucleosynthesis epoch.

In contradistinction to the cosmological model discussed in
\cite{RGTypeIa1,RGTypeIa2}, the functions (\ref{functionsz})
cannot be determined explicitly in an exact form. However, an
exact implicit function of the type $F(G(z;\nu),\rM(z);\nu)=0$ can
be derived after integrating the differential equations. For the
matter-dominated epoch the result reads
\begin{equation}\label{implicit}
\frac{1}{g(z;\nu)}-1+\nu\,\ln\left[\frac{1-\nu\,g(z;\nu)}{g(z;\nu)(1-\nu)}\right]=
\nu\,\ln\left[\frac{\rM(z)+\CC_0-\nu\,\rc^0}{\rM^0+\CC_0-\nu\,\rc^0}\right]\,,
\end{equation}
with
\begin{equation}\label{grho}
g(z;\nu)\equiv\frac{G(z;\nu)}{G_0}\,,\ \ \ \ \ \
\rM(z)=\rM^0(1+z)^3\,.
\end{equation}
The value of the gravitational constant and the matter density at
present ($z=0$) are $G_0$ and $\rM^0$ respectively, $\rc^0$ being
the critical density. We may check that (\ref{implicit}) is
satisfied by $g(0;\nu)=1$ for any $\nu$, as it could be expected.
The expression (\ref{implicit}) defines implicitly the function
$g=g(z;\nu)$, and once this is known the other cosmological
function $\CC=\CC(z;\nu)$ is obtained from
\begin{equation}\label{CCz}
\CC(z;\nu)=\frac{\CC_0+\nu\,\rM(z)\,g(z;\nu)-\nu\,\rc^0}{1-\nu\,\,g(z;\nu)}\,.
\end{equation}
Obviously this satisfies $\CC(0;\nu)=\CC_0$ for any $\nu$, as it
could also be expected.

Although the explicit analytic form cannot be obtained for
arbitrary $\nu$, it is possible to derive an explicit result for
small $\nu\ll 1$ within perturbation theory. This situation
actually adapts very well to the theoretical expectations on $\nu$
based on effective field theory
arguments\,\cite{JHEPCC1,RGTypeIa1}. However, there are also
phenomenological reasons, mainly from
nucleosynthesis\,\cite{RGTypeIa1,RGTypeIa2} and also from the CMB
and LSS\,\cite{Wang,Opher}, leading to the same
conclusion\,\footnote{For instance, in \cite{Wang} it is shown
that values of $\nu$ ($\epsilon/3$ in their notation) of order of
$\nu_0$ (Eq.\,(\ref{nu0})) or less are compatible with current
data on CMB and LSS.}. Expanding the exact formulae above up to
first order in $\nu$ we may compute the relative correction to the
gravitational constant:
\begin{equation}\label{gfirstnu}
\delta_G (z;\nu)\equiv \frac{G(z;\nu)-G_0}{G_0}=g(z;\nu)-1 \simeq
-\nu\,\ln\left[\frac{\rM(z)+\CC_0}{\rM^0+\CC_0}\right]=
-\nu\,\ln\left(\OM(z)+\OL^0\right)\,,
\end{equation}
where $\OM(z)\equiv\rM(z)/\rc^0$. Recalling that the present day
cosmological parameters, $\OM^0=\rM^0/\rc^0$ and
$\OL^0=\CC_0/\rc^0=\rho_{\CC}^0/\rc^0$, do fulfill the sum rule
$\OM^0+\OL^0=1$ for the flat space geometry,  we see that the
previous expression satisfies $\delta_G (0;\nu)=0$, as expected.
The above formula can also be obtained from Eq.\,(\ref{GH}) at
first order in $\nu$ upon replacing $H^2$ with the expression
dictated by Friedmann's equation (\ref{FL1}) with $k=0$. For
$\nu>0$ (resp. $\nu<0$) the gravitational constant was smaller
(resp. higher) in the past as compared to the present time. This
was already patent from Eq.\,(\ref{GH}) when we remarked that,
for $\nu>0$, $G$ becomes an asymptotically free coupling, which
means that at higher energies (i.e. when we look back to early
times) the coupling becomes smaller. Similarly, the explicit
result for $\Lambda(z;\nu)$ can be cast, to first order in $\nu$,
as follows:

\begin{figure}[t]
\centerline{\resizebox{1.0\textwidth}{!}{\includegraphics{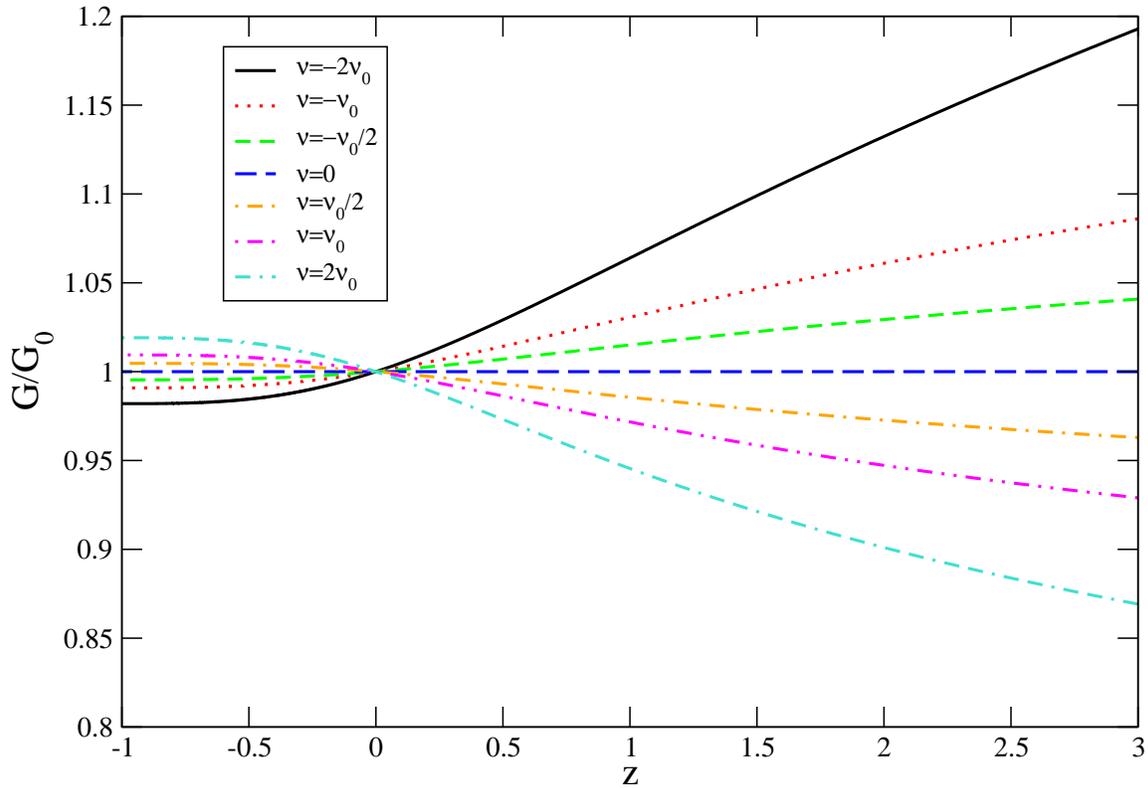}}}
\caption{\label{fig:g} The dependence of the gravitational
coupling constant $G=G(z;\nu)$ on the cosmological redshift $z$
for the six values (\protect\ref{row}) of the parameter $\nu$,
where $\nu_{0}$ is given by Eq.\,(\protect\ref{nu}). We assume a
flat cosmology, $k=0$, with $\Omega_M^0=0.3$, $\OL^0=0.7$. For the
parameter values of the order $\nu_{0}$, there is a significant
variation of $G$ at the redshifts amenable to high-z\, SNe Ia
observations.}
\end{figure}

\begin{figure}[t]
\centerline{\resizebox{1.0\textwidth}{!}{\includegraphics{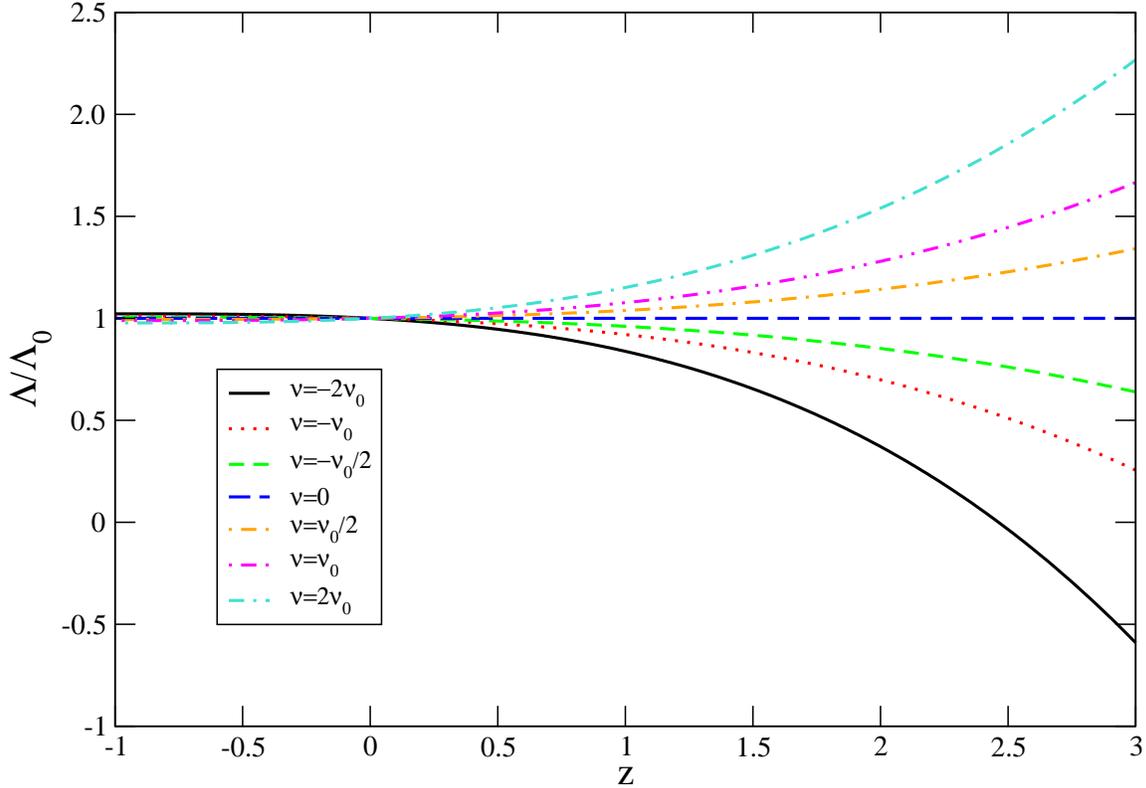}}}
\caption{\label{fig:xlam} The dependence of the cosmological term
energy density $\CC \equiv \rL(z;\nu)$ on the cosmological
redshift for values of the parameter $\nu$ as in
Fig.\,\protect\ref{fig:g}. For these parameter values there exists
significant variation  of $\CC$ (expressed in units of its present
value $\CC_0\equiv\rL^0$) which may be detected by the high-z SNIa
observations.}
\end{figure}

\begin{equation}\label{CCnu1}
\delta_{\CC}(z;\nu)\equiv\frac{\delta\CC(z;\nu)}
{\CC_0}=\frac{\CC(z;\nu)-\CC_0}{\CC_0}=
\nu\,\frac{\OM^0}{\OL^0}\,\left[(1+z)^3-1\right]\,.
\end{equation}
For convenience we have expressed the result in terms of the
relative deviation of $\CC(z;\nu)$ with respect to the present
value $\CC_0$, in order to better compare with
Ref.\,\cite{RGTypeIa1,RGTypeIa2}. Amazingly, to first order in
$\nu$, the cubic law (\ref{CCnu1}) is identical to the one
obtained in the latter references where a RG cosmological model
with variable $\CC$ and fixed gravitational constant was
considered. This means that the promising phenomenological
analysis made in \cite{RGTypeIa2} concerning the redshift
dependence of the cosmological constant remains essentially
intact. Let us, however, mention that the two models depart from
one another to second order in $\nu$. For the model under
consideration we find
\begin{equation}\label{CCnu2}
\delta_{\CC}(z;\nu)=
\frac{\OM^0}{\OL^0}\,\left[(1+z)^3-1\right](\nu+\nu^2)-
\left[1+\frac{\OM^0}{\OL^0}\,(1+z)^3\right]\,\ln\left[\OM^0\,(1+z)^3+\OL^0\right]\,\nu^2\,.
\end{equation}
To first order in $\nu$ we recover equation (\ref{CCnu1}), and
moreover we see that $\delta_{\CC}(0;\nu)=0$ for any $\nu$, as it
should be.

The previous (approximate) results are confirmed and refined
through a numerical analysis of the exact formulas
(\ref{implicit})-(\ref{CCz}). We take as representative values of
$\nu$ some multiples of the typical value $\nu_0$,
Eq.\,(\ref{nu0}). Specifically, we consider the six cases with
different signs represented by the $\nu$-data row:
\begin{equation}\label{row}
\overrightarrow{\nu}=\left(-2\nu_0,-\nu_0,-\frac{1}{2}\,\nu_0,
+\frac{1}{2}\nu_0,+\nu_0,+2\,\nu_0\right)\,.
\end{equation}
The results are presented in Fig.\,\ref{fig:g} and \ref{fig:xlam}
for the standard values of the cosmological
parameters\,\cite{R98P99}: $\Omega_M^0=0.3$, $\OL^0=0.7$.
Specifically, in Fig.\,\ref{fig:g} we plot the exact function
$G=G(z;\nu)$ for the six values of $\nu$ listed above. Similarly,
in Fig.\,\ref{fig:xlam} we plot the function $\CC=\CC(z;\nu)$ for
the same values of the $\nu$ parameter. We have verified these
results also by direct numerical solution of the differential
equations using \textit{Mathematica}\,\cite{Wolfram}. By looking
at Fig.\,\ref{fig:g} we may read off the relative correction
$\delta_{G}$ on $G$.  We see, for example, that significant
corrections $\delta_G\simeq -5\%$ are predicted at redshift
$z=1.5$, if $\nu\simeq\nu_0$. Furthermore, for the same values of
$\nu$ and $z$, we find $\delta_{\CC}\simeq 16\%$ (cf.
Fig.\,\ref{fig:xlam}). This correction to $\CC$ should be
perfectly measurable by SNAP\cite{SNAP}. Recall that this
experiment aims at exploring cosmological depths up to $z=1.7$,
and that other experiments are planning to go even beyond
(including the HST)\,\cite{GOODS}.

We should point out that cosmological corrections to $G$ of a few
percent at high redshifts are not in contradiction to present
bounds on the gravitational constant\,\cite{Hartle,Turyshev}. The
reason is that the existing bounds are ``local'', i.e. at the
scale of the Solar System, so that they do not provide any
information on possible variations of $G$ at cosmological
distances. Actually it is difficult to measure the correction
$\delta_G$ at the cosmological level, at least in the present
matter-dominated epoch. There is no direct way to measure $G$ at
distances of hundreds to thousands of Megaparsecs, unless we
re-interpret the observed cluster dynamics in terms of $G$
variation on a cosmological level, rather than in terms of CDM.
We will further elaborate on this kind of idea in the next
section, but restricting ourselves for the moment to the more
familiar galactic domain. On the other hand, if we extrapolate
back our running formula for $G$ to the radiation epoch we might
expect to find some additional information. Particularly relevant
is the nucleosynthesis epoch ($z\sim 10^9$). Performing the
corresponding change  $\rM\rightarrow\rR$ in Eq.\,(\ref{grho}),
with $\rR(z)=\rR^0\,(1+z)^4$, we find that for $\nu$ of order
$10^{-2}$ the relative variation $\delta_G$ lies within the limits
set by the latest analyses on $G$ variation at nucleosynthesis,
which give a margin of $\sim 40\%$ at the $95\%$
C.L.\,\cite{Gnucleosynth} Thus the existing nucleosynthesis limits
placed on $\nu$, based on considerations on $\CC$ variation
alone\,\cite{RGTypeIa1,RGTypeIa2}, are essentially unchanged.

\section{A possible application to flat rotation curves of galaxies}
\label{sect:Kepler}

In this section we speculate on the possibility to apply similar
RG methods to the more restricted astrophysical domains, such as
the galactic scales.  While we have seen that it is difficult to
measure the correction $\delta_G$ at the cosmological level, let
us explore the possibility to measure local variations of the
gravitational constant (induced by the RG) at the level of a
typical galaxy, and extract some potential consequences concerning
the origin of the flat rotation curves in its halo -- usually
associated to the presence of undetected
CDM\,\cite{Peebles}\,\footnote{We point out that previous attempts
exist in the literature to use the RG to explain the flat rotation
curves of galaxies\,\cite{Oldpapers}. However, these old papers
are based on the RGE of higher derivative quantum gravity. In
this theory the RGE's for individual $\CC$ and $G^{-1}$ are gauge
dependent and only the RGE for a dimensionless ratio is
unambiguous -- see e.g. \cite{book} and references therein.}.
Here again the main question is to identify the scale of the RG.
It can no longer be $\mu\simeq H$ because measurements within our
galaxy are not cosmological measurements, just astrophysical.
Indeed, $H$ does not change on the scale of a galaxy (a bound
system!). However, if $G$ varies according to some RGE, there
must be some new order parameter for the local RG scale. For
simplicity let us treat the galaxy as a central core surrounded
by  a spherically symmetric mass distribution. Then, from simple
dimensional arguments we may assume that the appropriate RG scale
for the galactic dynamics is given by
\begin{equation}\label{newscale}
\mu=\frac{\kappa}{r}\,,
\end{equation}
where $\kappa$ is some (dimensionless) constant and $r$ is the
radial distance (in the ``renormalized metric'', see below)
measured from the center of the galaxy.  If we assume that the
renormalization effects from the RG are small enough (after all
they cannot change dramatically the Newtonian picture), we may
take $r$ as the standard radial coordinate in classical celestial
mechanics. Notice that $\mu$ can never go smoothly to zero
because $r$ is bounded from above. This means that at the
galactic level we do not expect that the corresponding expansion
$G(\mu)^{-1}$ in (\ref{RGEInt}) is analytic near $\mu=0$ because
the system is intrinsically finite-sized. However, by
artificially extending the size of the local system we expect
that the same IR behavior of the cosmological case mentioned in
Section \ref{sect:RGmodel2} should be encountered. This provides
a connection between the two regimes (local and cosmological)
that can be used to establish the consistency of the RG
application to the local case. The Ansatz (\ref{newscale})
obviously implies that spatial gradients of $G$ (and may be also
of $\CC$) are allowed. It should be clear that this is not in
conflict with the Cosmological Principle because the latter
applies to the Universe in the large, not to local systems such
as galaxies or clusters of galaxies\,\footnote{The Cosmological
Principle indeed applies only to fundamental observers, and
therefore to extremely large scale systems at the level of
superclusters of galaxies\,\cite{Rindler}.}. Although the Ansatz
(\ref{newscale}) is plausible,  we may try to motivate it from
the point of view of the scale setting procedure
(\ref{rhoimplicit}). Strictly speaking, the latter has been
rigorously formulated only for the transition from the cosmic
time evolution to scale evolution. However, given the fact that
at very large scales $r$ is of order of the horizon, $1/H$, we
see that with the hypothesis (\ref{newscale}) the original scale
setting used in previous sections, $\mu\simeq H$, is consistently
met on cosmological scales. Consider the typical density profile
for flat rotation curves:
\begin{equation}\label{flatprofile}
\rho(r)=\frac{\rho_0}{1+(r/r_0)^2}\,.
\end{equation}
Here $r_0$  is a reference galactic length of the visible part of
the galaxy; e.g. in the case of our galaxy it can be taken as the
``local'' position (at the galactic scale) of our Solar System,
where the gravitational constant is supposed to be $G_0$
essentially everywhere,  and $\rho_0$ is the average density of
the galaxy well before reaching the halo regime. Clearly
$\rho(r)\rightarrow\rho_0$ for $r\ll r_0$ whilst $\rho(r)$ falls
off as $1/r^2$ for $r\gg r_0$ (i.e. deep in the halo). This
behavior is exactly what is needed to have a linear growth of the
galactic mass deep in the halo region,
\begin{equation}\label{gmass}
{\cal M}(r)=\int_{r\gg r_0} d^3r\ \rho(r) \propto r\,,
\end{equation}
and therefore a constant rotation velocity in it:
\begin{equation}\label{flatveloc}
v=\sqrt{\frac{G\,{\cal M}(r)}{r}}\rightarrow {\rm const.}
\end{equation}
This is the classical ``explanation'' for the flat rotation
curves, once the galactic density profile (\ref{flatprofile}) is
accepted by fiat. One possibility (the most commonly accepted one)
is that (\ref{flatprofile}) is associated to the presence of CDM
distributed in the halo region $r\gg r_0$. However, an alternative
justification for this density function (and perhaps also the
ultimate origin of the flat rotation curves) could actually be
linked to the RG running of the gravitational constant as a
function of the new scale parameter (\ref{newscale}). A hint that
this could be the case is obtained by  substituting the density
(\ref{flatprofile}) into $\rho$ in Eq.\,(\ref{muexample1}). It is
easy to see that when the halo regime sets in, the resulting
formula behaves precisely in the form (\ref{newscale}). We will
see below that the density (\ref{flatprofile}) although it is not
the real mass density of the galaxy in our framework (because we
do \textit{not} assume a priori the existence of CDM), it can be
reinterpreted as the ``renormalized'' mass density, i.e. the
``effective mass density'' of the galaxy after considering the
renormalization effects on Newton's gravitational constant. In
this sense it is this expression that must be used to determine
the RG scale. Let us therefore take these features as heuristic
motivations to adopt $\mu$ from (\ref{newscale}) as a valid RG
scale for the galactic regime, and let us explore the ensuing
consequences within our RG framework. Upon replacing $\mu=H$ with
$\mu=\kappa/r$ and $H_0$ with $\kappa/r_0$ in our running formula
(\ref{GH}) we find
\begin{equation}\label{GH2}
G(r)=\frac{G_0}{1+\nu\,\ln\left(r_0^2/r^2\right)}\,.
\end{equation}
We assume $\nu>0$ in (\ref{GH2}) so that $G$ is asymptotically
free, and therefore it increases with $r$. This kind of growing
behaviour of $G$ with the distance is reminiscent of the above
mentioned possibility that gravity can lead to non-perturbative
effects in the far IR. The running of $G(r)$ in the surroundings
of the Solar System (i.e. for $r\simeq r_0$) is very small; in
practice there is no running at all, for the entire Solar System
is represented by the point $r=r_0$ without extension. This
insures that we normalize our $G_0$ above with the value obtained
from highly accurate measurements of $G$ performed in our
neighborhood\,\cite{Hartle,Turyshev}. Any departure from this
value was never accessible to us because we never measured $G$
(at least directly and accurately) outside the Solar System.
Moreover, the formula above cannot be applied for $r$ near $r=0$.
The reason is that the geometry is governed by the Schwarzschild
metric (see below), and therefore there is an event horizon at
some point $r=2\,G{\cal M}\ll r_0$, where ${\cal M}$ is the mass
of the galaxy concentrated at its center. We can only assume
valid Eq.\,(\ref{GH2}) well beyond this limit.  Next we consider
the modified Einstein-Hilbert action with variable $G$:
\begin{eqnarray}
S^{(1)}_{EH} = -\int d^4 x\sqrt{-g}\,. \frac{R}{16\pi\,G(r)}\,.
\label{modHE}
\end{eqnarray}
If G were constant, we would expect that the metric that solves
the galactic dynamics at large $r$ is the spherically symmetric
Schwarzschild metric. For variable $G$, however, the situation is
more complicated. One could vary the action (\ref{modHE}) and try
to find out the field equations, but it is hard to solve them
exactly. Whatever the solution is, at least we know that the
spherical symmetry must still be there. Therefore, in the absence
of an exact result we may use the fact that $G(r)$ is given by
(\ref{GH2}) and that $\nu$ is a small parameter. Following the
conformal symmetry methods of \cite{shocom,PSW}, we introduce the
auxiliary action
\begin{eqnarray}
S^{(2)}_{EH} = -\frac{1}{16\pi G_0}\,\int d^4 x\sqrt{-\bar{g}}\,
\,\bar{R} \,, \label{EHm}
\end{eqnarray}
with respect to the auxiliary metric $\bar{g}_{\mu\nu}$, and with
\textit{constant} $G=G_0$. This action transforms into the action
\begin{eqnarray}
S^{(3)}_{EH} = -\frac{1}{16\pi G_0}\,\int d^4 x\sqrt{-g}\,
\,e^{2\,s}\left[\,R + 6\,(\partial s)^2\,\right]\,. \label{cosmon}
\end{eqnarray}
under the local conformal transformation
\begin{equation}\label{conformal}
\bar{g}_{\mu\nu}=e^{2s}\,{g}_{\mu\nu}\,,
\end{equation}
where $s=s(r)$ is a function of $r$. With this trick, let us now
choose the following conformal factor:
\begin{equation}\label{cf}
e^{2s}=\frac{G_0}{G(r)}={1+\nu\,\ln\left(r_0^2/r^2\right)}\,,
\end{equation}
where $G(r)$ is given by (\ref{GH2}). Then the action
(\ref{cosmon}) becomes the $r$-dependent G action (\ref{modHE})
except for the extra derivative terms $(\partial s)^2$:
\begin{eqnarray}
S^{(4)}_{EH} = -\int d^4 x\sqrt{-g}\, \frac{R}{16\pi\,G(r)}-6\int
d^4 x\sqrt{-g}\,\,\frac{(\partial s)^2}{16\pi\,G(r)}\,.
\label{modHE2}
\end{eqnarray}
Notice that the square derivative terms are of higher order in
$\nu$ with respect to the Lagrangian density $\sqrt{-g}\,R/G(r)$,
and can be neglected. Indeed, differentiating (\ref{cf}) on both
sides gives $e^{2s}\partial s/\partial r =-\nu/r$ and hence the
square derivative that appears as the extra term in
(\ref{modHE2}) becomes of order $(\partial s)^2\sim \nu^2/r^2$.
For a typical radius $r$ comparable to the galactic size we can
fully neglect this term as compared to $R\sim 1/r^2$ because
contributions of order $\nu^2$ are very small. It follows that the
solution of the field equations associated to the original action
(\ref{modHE}) with variable $G$ is, to first order in $\nu$,
\begin{equation}\label{conformal2}
g_{\mu\nu}=e^{-2s}\,\bar{g}_{\mu\nu}=
\left(1+2\nu\ln\frac{r}{r_0}\right)\,\bar{g}_{\mu\nu}\,.
\end{equation}
From the arguments above, $\bar{g}_{\mu\nu}$ must be the solution
of the ordinary Einstein equations with constant $G_0$
corresponding to the galaxy system treated as a point-like core
of mass ${\cal M}$. Therefore $\bar{g}_{\mu\nu}$ on the
\textit{r.h.s.} of (\ref{conformal2}) must necessarily be the
standard Schwarzschild metric corresponding to constant $G_0$:
\begin{equation}\label{Schwarz0}
ds^2_{\rm Schw}=\left(1-\frac{2\,G_0\,{\cal M}}{r}\right)\,dt^2-
\frac{dr^2}{1-2\,G_0\,{\cal M}/r}-r^2\,d\Omega^2\,.
\end{equation}
From (\ref{conformal2}) we learn that the solution of the field
equations for $G=G(r)$ must be a modified (or ``renormalized'')
Schwarzschild's metric. To first order in $\nu$ the
renormalization amounts to the new line element
\begin{equation}\label{Schwarz1}
ds^2=\left(1+2\nu\ln\frac{r}{r_0}\right)\,ds^2_{\rm Schw}\,.
\end{equation}
Let us now take the $g_{00}$ component of this modified metric
and recall that the classic potential $\Phi$ is identified from
$g_{00}=1+2\Phi$. To first order in $\nu$ it reads
\begin{equation}\label{NNP}
\Phi(r)=-G_{0}\,\frac{\cal
M}{r}+\nu\,\left(1-2\,G_{0}\,\frac{\cal M}{r}\right)\,\ln
\frac{r}{r_0}\,.
\end{equation}
The corresponding force per unit mass becomes
\begin{equation}\label{NNF}
F(r)=-\frac{d\Phi}{dr}=-(1-2\,\nu)\,G_{0}\,\frac{\cal
M}{r^2}-\,\nu\,\left(\frac{1}{r}+\,G_{0}\,\frac{\cal M}{r^2}\,\ln
\frac{r^2}{r^2_0}\right)\,.
\end{equation}
This intriguing result shows that for sufficiently large $r$
(namely, in the halo of the galaxy where $r>>r_0$) the force that
we have found is not purely Newtonian as it does no longer decay
as $1/r^2$ at high distances, but as $1/r$.  The coefficient of
the $1/r$ term is precisely the parameter $\nu$ of our
renormalization group model. Equating this force to the
centripetal force $-v^2/r$ and neglecting for the moment the
$1/r^2$ (Newton's) term and the $(\ln r)/r^2$ term, one finds
that for large $r$ the rotation velocity is virtually constant,
i.e. it does not depend on $r$ (the rotation curve becomes flat).
Such asymptotic velocity is thus determined in good approximation
by the square root of our original parameter $v=\sqrt\nu$. In
conventional units the velocity reads $v/c=\sqrt\nu\ $. Of course
we expect $v/c<<1$ for any rotating object in the halo. But this
can be well accomplished through the coefficient $\nu$, which was
always expected small in our effective field theory approach to
the $\beta$-function of $\CC$ down the Planck scale, see
(\ref{nu}). We argued that $\nu$ could typically be of order
$10^{-2}$ for the cosmological considerations. However, we may
ask now: how much small must it be to describe the physics of the
flat rotation curves of galaxies? The rotation velocities in most
galaxies are of a few hundred Km/s, say $200-300$
Km/s\,\cite{Peebles}. It follows that if this model is to
describe them, $\nu$ should be of order
\begin{equation}\label{nuvalue}
\nu\sim 10^{-6}\,.
\end{equation}
Obviously a very small value! It implies that all of the
approximations that we have made for $\nu\ll 1$ should be
perfectly acceptable. And of course, from the cosmological point
of view, it is neatly compatible with the CMB and LSS measurements
\,\cite{Opher}. With this value of $\nu$ we may explicitly check
that the dominance of the leading extra term  $\nu/r$ in
(\ref{NNF}) over the Newtonian one, $G_0\,{\cal M}/r^2$, really
takes place in the halo of the galaxy, where $r>>r_0$. Indeed,
for a typical galaxy with a mass $10^{10-11}$ times the mass
$M_{\odot}$ of our Sun, and with the $\nu$ value (\ref{nuvalue})
obtained in our RG framework, it is easy to check that the
$\nu/r$ term is comparable to the Newtonian force just around a
scale of several to ten $Kpc$, namely near the onset of the halo
region. Beyond $10\,Kpc$, and of course deep in the halo region,
the ``$\nu$-force'' is dominant.

If we keep the other terms in (\ref{NNF}), except for the very
small correction to the Newtonian part of the force, we find the
following formula for the flat rotation velocity:
\begin{equation}\label{vc}
v/c=\sqrt{\nu+\,G_{0}\,\frac{\cal M}{r}\left(1+\nu\,\ln
\frac{r^2}{r^2_0}\right)}\simeq \sqrt{\nu+\,G_{0}\,\frac{\cal
M}{r}} \,,
\end{equation}
where in view of the small value (\ref{nuvalue}) we have neglected
$\nu\,\ln (r^2/r^2_0)\ll 1$. The previous result reads essentially
$v/c=\sqrt\nu$ up to a small correction which depends on the
total mass of the galaxy and the radial distance to the rotating
object in the halo. This is again an interesting feature, because
the velocity of the flat rotation curves is well-known to show
some departure from universality\,\cite{Peebles}. In fact, we can
further shape this formula such that it becomes closer to the
observed behavior. From the modified Newton's law we may estimate
the effective mass (or ``renormalized mass ${\cal M}_{r}$'') of
the galaxy associated to the renormalization of $G$. By equating
the purely Newtonian term in (\ref{NNF}) to the leading correction
proportional to $\nu$ we immediately conclude that the
``effective additional mass'' of the galaxy is proportional to
$r$ in the halo region,
\begin{equation}\label{deltaM}
 \delta{\cal M}(r;\nu)\simeq \left(\frac{\nu}{G_0}\right)\,\,r\ \ \ \  (r\gg
r_0)\,.
\end{equation}
The total renormalized mass of the galaxy (the one which is
actually observed, if one adopts the purely Newtonian approach)
is thus given by the ``bare mass'' (i.e. the ``real'' or ordinary
mass ${\cal M}$ made out of baryons) plus the renormalization
effect computed above: ${\cal M}_{r}={\cal M}+\delta{\cal M}$. It
should, however, be clear that this is a huge renormalization
effect, because $\delta{\cal M}$ is comparable to (actually
larger than) ${\cal M}$ as it could be expected from the fact
that, in the language of the CDM, the extra amount of matter in a
typical galaxy is known to be dominant over the luminous one:
roughly ten times the total baryonic matter (whether luminous or
not). It means that we should have $\delta{\cal M}\gg {\cal M}$,
equivalently ${\cal M}_{r}\simeq \delta{\cal M}$. This feature is
nicely realized in our RG framework and again emphasizes the
potential non-perturbative character of the gravitational effects
in the IR regime, similar to the strong low-energy dynamics of
QCD. To see this quantitatively, let us take $r$ deep in the halo
region, say in the range $10-50\,Kpc.$, and let us assume a
galaxy with a physical mass ${\cal M}=10^{10}\,M_{\odot}$
(typically the case of the total baryonic mass of our galaxy). It
is easy to estimate from (\ref{deltaM}), using the value of $\nu$
given in (\ref{nuvalue}), that $\delta{\cal M}\simeq 10\,{\cal
M}$, which is in the right order of magnitude to describe a total
effective mass around ${\cal M}_{r}\sim 10^{11}\,M_{\odot}$
without invoking the existence of CDM. Strictly speaking, there
is no physical halo in the RG approach, but at some point the
local RG description must break down. This should define an
``effective RG halo'' of the galaxy, beyond which  a larger scale
RG picture should take over.

The linear growth of the mass given in Eq.\, (\ref{deltaM})
implies that the ``effective density'' of the galaxy in the halo
region behaves as
\begin{equation}\label{rhoeff}
 \rho(r)\simeq
(\frac{\nu}{4\,\pi\,G_0})\,\left(\frac{1}{r^2}\right) \ \ \  (r\gg
r_0)\,.
\end{equation}
This effective density replaces the material effects associated to
the CDM matter density (\ref{flatprofile}). Comparing both
densities in the region $r\gg r_0$ we obtain
$r_0^2\,\rho_0=\nu/4\pi G_0$. Setting $\rho_0\sim {\cal M}/r_0^3$
we have ${\cal M}/r_0\sim \nu/G_0$, which is indeed satisfied
within order of magnitude for a similar set of inputs as before.
Finally, since the renormalized mass of the galaxy is
proportional to the galactic radius in the halo region, we
conclude that the rotation velocity (\ref{vc}) can be cast as
\begin{equation}\label{vc2}
v(r;{\cal M})/c\simeq \sqrt{\nu}\,\left[1+\varepsilon (r;{\cal
M})\right]^{1/2}\ \ \ \  (r\gg r_0)\,,
\end{equation}
where  $\varepsilon (r;{\cal M})$ is a small correction that
exhibits the departure of the flat rotation curves from
universality. It reads
\begin{equation}\label{vepsilon}
\varepsilon (r;{\cal M})=\frac{{\cal M}}{{\cal M}_{\nu}(r)}\,,
\end{equation}
with ${\cal M}_{\nu}(r)\equiv\delta {\cal M}(r;\nu)$ given by
Eq.\,(\ref{deltaM}). Notice that ${\cal M}_{\nu}(r)$ does {\em
not} depend on the mass of the galaxy, but only on $\nu$ and on
the radial position of the rotating object. We have seen above
that typically ${\cal M}_{\nu}(r)\sim 10\,{\cal M}$ for $r\gg
r_0$. Therefore, we conclude that the flat rotation velocity
becomes a function $v=v(r;{\cal M})$. In other words, the small
corrections to universality depend on the mass of the galaxy and
of the radial location of the object, another observed feature.
Only further refinements of the RG approach and the galaxy model
may perhaps shed some light on the detailed numerical predictions
concerning the value of the rotation velocity and its dispersion.
Our formula above can easily accommodate velocity dispersions of
order $10\%$. However, we cannot expect perfection. Recall that
we modeled the galaxy as a spherical distribution of matter while
it is roughly disk-shaped, thus other geometric and dimensionful
variables might enter (including the mass ${\cal M}$ of the
galaxy) to refine our scale Ansatz (\ref{newscale}). This could
have some further impact in the departure of the flat rotation
velocity from universality.

We should recall that the flat rotation curves of the galaxies
can be fitted within the Modified Newtonian Dynamics
(MOND)\,\cite{MOND}. This phenomenological model requires
gravitation to depart from the standard Newtonian theory in the
extra-galactic regime ($r\gg r_0$). The MOND potential takes the
form
\begin{equation}\label{MOND}
\tilde{\Phi}(r)=-G_{0}\,\frac{\cal M}{r}+\sqrt{a_0\,G\,{\cal M}}\
\ln \frac{r}{r_0}\,.
\end{equation}
The structure of the extra-galactic modification (the second term)
is such that one can fit both the flat rotation curves and the
departure of these curves from strict universality. The first
feature is achieved from the logarithmic form of the correction,
and the second feature is implemented by allowing for the
coefficient of the logarithm to depend on the product $G\,{\cal
M}$, where ${\cal M}$ is the mass of the galaxy. Obviously the
overall coefficient must be dimensionless, and therefore one is
forced to introduce a new ``fundamental acceleration'' parameter
$a_0$ in Eq.\,(\ref{MOND}), which can be fitted observationally to
a value of order $10^{-8}\,cm\,sec^{-2}$. The combination of these
two features is all what is needed to simultaneously account for
the leading and subleading observational effects of the galaxy
rotation curves without invoking the existence of CDM haloes: to
wit, the flatness of these rotation curves and the so-called
Tully-Fisher's law, $v^4\propto{\cal L}$, where $v$ is the
asymptotic value of the rotation velocity and ${\cal L}$ is the
absolute luminosity of the galaxy\,\cite{TullyFisher}. This law
expresses some departure of the flat rotation curves from
universality and can be rephrased in terms of the mass of the
galaxy. If one makes the standard assumption that the
mass-to-light ratio, ${\cal M}/{\cal L}$, is essentially constant
for spiral galaxies, then $v^4\propto {\cal M}$. MOND indeed
implements this relation in the form $v^4=a_0\,G\,{\cal M}$ -- as
it trivially follows from (\ref{MOND}). However, it must be
clearly stated that in spite of the great phenomenological
success of MOND, so far none of the relativistic gravitational
theories proposed up to date (aiming to underpin that
phenomenological model from first principles) can be considered as
free of theoretical and experimental inconsistencies, including
the recent attempt in \,\cite{Bekenstein}. In this sense, it is a
welcome feature that we can derive the leading effect (the
existence of the flat rotation curves themselves) from our RG
framework. This is obvious when we compare our modified Newtonian
potential (\ref{NNP}) and the MOND potential (\ref{MOND}). The
two potentials have in common the necessary logarithmic term to
account for the leading correction to Kepler's law entailing the
flat rotation feature. However, the RG-inspired potential
(\ref{NNP}) does not lead to the (subleading) Tully-Fisher's
effect, while the MOND potential contains the piece
$\sqrt{a_0\,G\,{\cal M}}$ just ordered \textit{ad hoc} to
reproduce this additional feature.  It could be that the
RG-potential misses this piece due to the simplifications we have
made to reach the leading result. We hope that further
investigations will help to clarify this issue. In the meanwhile
Eq.\,(\ref{vepsilon}) above already suggests that some
mass-dependent departure from universality is predicted in the
simplest RG approach.

With  $\nu$ as small as (\ref{nuvalue}) the increase of $G(r)$
from (\ref{GH2}) remains negligible, even at cosmological
distances. Therefore, it is conceivable that $G(r)$ essentially
behaves as a constant, with no further significant variation up
to the horizon. At this point we recall that the motivation for
(\ref{GH2}) was, in contrast to the more formal one of
(\ref{GH}), essentially heuristic. However, since at very large
scales $r$ is of order of the horizon, $1/H$, the function
$G=G(r)$ merges asymptotically with $G=G(H)$. In this way the
original scale setting $\mu=H$ is retrieved on cosmological
scales and there is a  matching of the two scale settings in the
asymptotic regime. Notwithstanding, we should clarify that we do
not expect a sharp transition from one RG regime to the other; we
rather expect the existence of some interpolating RG description
that connects scales of order $100\,Kpc$ (which characterizes the
maximum expected size of the galactic haloes) with scales of order
$100\,Mpc$ (above which the large scale cosmological picture
takes over). In the crossover region the effective density
(\ref{rhoeff}) should fall off faster than $1/r^2$.
Unfortunately, it is not possible to be more quantitative here on
the specific form of this transition, but we do expect it should
be there. After going through this transition the running formula
(\ref{GH2}) becomes (\ref{GH}) and the common, and essentially
constant, value of $G$ in the two scaling laws represents the
cosmological value of the gravitational constant. The only
measurable effect left in this scenario would be the
aforementioned correction to Newtonian celestial mechanics, which
could have an impact on dark matter issues at the galactic and
even at larger (cluster) scales. But we are not going to push
further this possibility here nor to contemplate the possibility
of other refinements. In particular, in the above considerations
we have ignored the potential running effects from the
cosmological constant at the galactic scale, which were assumed
to be small. We close this section by noticing that with $\nu$ of
the order of (\ref{nuvalue}), the GUT scale potentially
associated to these quantum effects would be $M_X\sim
10^{16}\,GeV$ rather than $10^{17}\,GeV$ (see Eq.\,(\ref{MX})),
therefore closer to the lowest expected SUSY-GUT scale.

\section{Discussion and conclusions}
\label{sect:conclusions}

In this paper we have further elaborated on the formulation of the
renormalization group cosmologies along the lines of
\,\cite{JHEPCC1, cosm,
Babic1,RGTypeIa1,RGTypeIa2,IRGA03,AHEP03,Babic2}. The main aim of
these RG cosmologies is to provide a fundamental answer, within
QFT, to the meaning of the most basic cosmological parameters
such as $H_0$, $G$, $\rM^0$ and $\CC_0\equiv\rL^0$, and at the
same time to try to understand their presently measured values
and possible interrelations among them. In doing so we have found
that the classical FLRW cosmologies become ``renormalization
group improved'', and therefore they become slightly modified by
quantum corrections associated to renormalization group equations
that govern the evolution of the parameters $\CC$ and $G$. This
evolution behaviour, primarily a scaling one within the framework
of the RG,  can be inter-converted into a temporal evolution
(i.e. in terms of the cosmic time or, if desired, in terms of the
redshift), after an appropriate identification of the scale
parameter $\mu$ of the RG. We have found that $\mu\sim H(t)$ is a
consistent identification for the class of RG cosmologies in
which $\CC$ evolves quadratically, and $G$ evolves
logarithmically, with $\mu$. And we have shown that this
combination of running laws is a most promising one from the
point of view of phenomenology. The scaling (and redshift)
variation of these two laws is very simple and it is controlled
by a single parameter, $\nu$ -- cf. Eq.\,(\ref{nu}). For $\nu>0$
the running law for the gravitational constant exhibits
(logarithmic) asymptotic freedom, much as the strong coupling
constant in QCD. For $\nu\ll 1$ the quadratic evolution law for
$\CC$ turns out to be very similar to the one previously found in
\cite{RGTypeIa1,RGTypeIa2} and therefore it can be tested in the
future experiments by SNAP\,\cite{SNAP}, provided $\nu$ is not
much smaller than $\nu_0\sim 10^{-2}$. This value is presently
consistent with data from high-z SNe Ia, CMB and LSS.
Furthermore, from the point of view of the cosmological constant
problem, we have seen that this RG framework can shed some light
to the coincidence cosmological problem ($\rL^0\simeq\rM^0$ ?)
because it provides a natural explanation of the puzzle of why
the energy scale associated to the cosmological constant is at
present the geometric mean of the Hubble parameter and the Planck
mass: $E_{\CC}(t_0)\simeq\sqrt{M_P\,H_0}$.

It is important to remark that the quadratic law (\ref{CCH})
associated to our RG framework is not just a proportionality law
$\CC\sim H^2$, but a variation law: $\delta \CC\sim H^2$. In
contradistinction to the former, the latter actually introduces
an additive term. This is because it originally comes from a RGE
of the form $d\CC/d\ln H\propto H^2$, which has to be integrated,
see Eq.\,(\ref{RG01}). Therefore the ensuing cosmologies are very
different from the $\CC\sim H^2$ ones. Only the latter have been
amply discussed in the past in many places of the
literature\,\cite{CCphenom,Overduin}, including some recently
raised criticisms\,\cite{Wang,Hsu}. In contrast the $\delta
\CC\sim H^2$ law is, to the best of our knowledge, relatively new
in the literature\,\cite{JHEPCC1}, and its motivation within QFT
gives it a credit at a much more fundamental level. Let us
mention that some variation of this law has been proposed very
recently (on pure phenomenological grounds) in \cite{Wang}, in
the form $d\CC/da\propto dR/da$, where $R$ is the curvature
scalar.  However, since $\mu=R^{1/2}$ is, as mentioned above (and
in \cite{JHEPCC1}), a covariant generalization of $\mu=H$, it is
essentially the same law and so the conclusions are similar.
Therefore the RG cosmologies associated to $\delta \CC\sim H^2$
can also be formulated covariantly, without committing to the
particular FLRW metric. But from the practical point of view (and
also to better assess its physical meaning and potential
applications\,\cite{RGTypeIa2,AHEP03}) it is better to formulate
this law  making use of the Hubble parameter, instead of
curvature.

The last part of our paper is more speculative. We have tantalized
the possibility of extending the RG formalism to local
cosmological domains such as the galactic level. After arguing
that the local RG scale could naturally be $\mu\sim 1/r$, we have
solved the modified Schwarzschild metric in the limit $\nu\ll 1$
and derived a RG correction to Kepler's third law of celestial
mechanics. We have hypothesized that this quantum correction could
furnish an explanation for the flat rotation curves of the
galaxies, provided $\nu\sim 10^{-6}$. Obviously, this possibility
would exclude measurable effects on the cosmology side, where a
larger $\nu$ is needed, but it might hit square on one of the
biggest astrophysical conundrums of the last thirty years. This
result is curious enough and rather intriguing, and we have found
interesting to include it as an additional potential application
of our RG framework.  All in all, we deem this last part of our
work more speculative than the rest, because we have had to
descend from the cosmological level to the (more complicated)
local domain where cosmology is entangled with astrophysics, and
other factors may come into play.  Thus, more detailed theoretical
considerations are needed to formally establish the RG effects at
the galactic level,  which could be modified by other factors such
as the modeling of the galaxy and the details of the RG
framework. For this reason, even if $\nu$ is of order of $\nu_0$,
and therefore much larger than (\ref{nuvalue}), the corresponding
RG predictions at the galactic level are not necessary in
contradiction with the experimental situation. The actual effects
could be much smaller due to, say, the need to introduce some
significant modification in the Ansatz (\ref{newscale}), as noted
in the last section. Finally, even if accepting that $\nu$ is
small enough for the RG to be able to provide some explanation of
the flat rotation curves, it does not preclude the possibility
that galactic dark matter is present in the haloes of the
galaxies. It only shows that other, even more subtle, effects
could creep in. Whether or not these effects are directly related
to the RG must be further investigated, but already at the pure
phenomenological level the possibility that a logarithmic law of
the type (\ref{GH2}) could be behind the flat rotation curves
looks rather enticing. Obviously, even if it is not the sole
explanation, it may help to reduce the total amount of dark
matter needed to account for the flat rotation curves and also to
explain some features related to the deviation from universality
of the flat velocities. In any case it is clear that further
experimental work will also be necessary to decide on the
ultimate origin of the rotation curves of the
galaxies\,\cite{Prada}. In general, even if letting aside the
possibility of these local effects, our cosmological RG framework
provides a sound alternative to explain the nature of the
cosmological parameters and their potential connection to quantum
field theory. It also offers a link between the low-energy
physics characteristic of large scale Cosmology with the
high-energy interactions of Particle Physics. Last, but not
least, it constitutes a ready testable framework for the next
generation of cosmological experiments.
\\

{\bf Note Added:}  Shortly after the e-print Archive version of
this paper was submitted as  \texttt{hep-ph/0410095}, some other
works were submitted to the Archive -- see
\cite{ReuterWeyer,Moffat}-- which overlap significantly with the
content of Section 7. These works also use the RG to study the
flat rotation curves of the galaxies and arrive to conclusions
very similar to the ones presented here. Since, however, their
approach is different from ours we think it can be useful for the
reader if we comment on the similarities and also the main
differences between the two methods.

First of all we recall that two Renormalization Group formulations
for gravity are present in the literature, namely the perturbative
RG for curved space-time and the Wilsonian non-perturbative RG for
quantum gravity. The present work is based on the first of these
RG approaches (see \cite{birdav,book} for an introduction and
references). The object of our interest is the quantum theory of
matter fields on classical curved background. It is very
important that this theory is renormalizable and therefore
application of the perturbative RG to this case is a consistent
procedure. Unfortunately, the consistency does not mean that this
approach does not meet real difficulties. The existing
well-established formalism is essentially restricted by the
$\overline{\mbox{MS}}$-scheme of renormalization. Within this
approach one can learn how the observables dependent on the
dimensional parameter $\,\mu$, but the physical interpretation of
this parameter always requires a special additional effort.
Furthermore, the $\overline{\mbox{MS}}$-scheme is reliable only
in the UV limit and is not appropriate for the investigation of
the relatively weak IR scaling, that is for the study of
decoupling. An alternative mass-dependent scheme has been
recently applied in curved space-time, but only in the particular
case of linearized gravity \cite{apco}. In this case we can
observe the Appelquist\&Carazzone decoupling theorem\,\cite{AC}
to hold in the higher derivative sectors of the gravitational
action. The present and previous
\cite{JHEPCC1,Babic1,RGTypeIa1,RGTypeIa2} works are based on the
phenomenological assumption that a similar kind of decoupling
behaviour takes place for the cosmological constant. One can find
an extended discussion of this issue in the Appendix A of
\cite{RGTypeIa2}.

On the other hand, the alternative works \cite{ReuterWeyer}
(exactly as the previous ones \cite{Reuter03a,Bentivegna}, see
further references therein) deal with the IR effects of quantum
gravity. It is well known that quantum gravity is inconsistent
within the perturbative approach and hence the appropriate
version of the RG is necessarily non-perturbative.
Ref.\,\cite{ReuterWeyer} is based on a simplified version of the
Wilsonian approach to this non-perturbative RG framework. The
objective is the derivation of the functional of the flow
effective action, depending on the cut-off parameter $\,k$. This
parameter can be viewed as a direct analog of the parameter
$\,\mu\,$ for the perturbative RG in the
$\overline{\mbox{MS}}$-scheme. The main similarity between the
two parameters is that one also needs to properly identify
$\,k\,$ with certain physical quantity (e.g. the energy of the
gravitational quanta or the inverse of the cosmic time $\,t$).
Without such identification the theory has no physical output.
The choice of the scaling parameter in both RG approaches is not
constrained by the RG formalism itself, and in certain
(sufficiently simple) situations it can be guessed on the basis
of physical (e.g. dimensional) arguments. This is the case when
the system has just one relevant physical scale. For instance, in
the astrophysical case under consideration in Section
\ref{sect:Kepler} the galaxy is assumed to have spherical symmetry
and then the single relevant physical scale is the radial distance
$r$ from its center. This leads to the successful scale choice
(\ref{newscale}). On the other hand, in the cosmological context,
a possible way of constraining the scale setting is the procedure
applied in Section \ref{sect:RGmodel3}. This leads immediately to
our choice $\mu=H$ -- cf. Eq.\,(\ref{muH}) -- roughly consistent
with the scale setting $k\sim 1/t$ advocated in the alternative
RG approach of \cite{ReuterWeyer}. However, even this regular
procedure does not obviously provide universal solution for all
physical situations (for example, in multiple scale systems or in
the transition from the cosmological to the astrophysical
domain). Therefore, the link between the RG running and the
$\,r$-dependence in the astrophysical setting is an essentially
phenomenological assumption for both RG formulations, and needs
further investigation.

As we already mentioned above, in our RG approach we assume the
standard form of decoupling for the cosmological constant.
Similarly, the alternative RG approach -- the Wilsonian
non-perturbative formulation of the RG for quantum gravity --
requires also of strong phenomenological input. In the Wilsonian
version of the RG used in \cite{ReuterWeyer} the effective action
depends on infinitely many couplings associated to powers of
curvature and non-local invariants. Solving the exact functional
RG flow equation is virtually impossible at present and in
practice one truncates this effective action into a linear
combination of powers of curvature, the simplest possibility
being the powers one and zero, i.e. the Einstein-Hilbert term and
the cosmological term. The coefficients of these terms define the
flow trajectory $(G(k),\Lambda(k))$ projected onto this simplified
finite-dimensional space.

However, by virtue of this simplification the flow trajectory
$G=G(k)\,,\Lambda=\Lambda(k)$ cannot be derived from the RG flow
equation, and therefore an Ansatz on the trajectory is to be made
in general. For the galactic problem at hand, one typically
dismisses the role played by the cosmological term and is left
with proposing an (\textit{ad hoc}) modeling for the $G$ flow.
The model trajectory explored in \,\cite{ReuterWeyer} reads
$G(k)=1/k^q$ (with $q$ an arbitrary parameter). In order to
describe the flat rotation curves the authors of
\cite{ReuterWeyer} have to take for $q$ a very small value of
order $10^{-6}$. In contrast, in a previous
paper\,\cite{Bentivegna} it was assumed the particular value
$q=2$ in order to describe a RG cosmology with a (hypothetical)
IR fixed point. Obviously this trial and error strategy shows
that the state of the art in the RG flow effective action
approach is not good enough at present as to allow for
determining the flow trajectory $G=G(k)$ in a fully dynamical
way. It is nevertheless remarkable that after accepting the value
$10^{-6}$ the previous power law trajectory becomes effectively a
logarithmic running law, which is entirely similar to the one we
have obtained in Eq.\,(\ref{GH2}) on a very different basis.

Comparing the merits of the two RG approaches we can indicate two
advantages (from our point of view) of the one considered in the
present paper. {\it First}, we are dealing with the quantum
effects of matter fields, which definitely take place. The
necessary phenomenological implication concerns only the
technical (albeit serious) difficulty of evaluating the vacuum
effective action of massive fields on a curved background. In
contrast, the relevance of the quantum gravity effects and the
possibility of their description without solving the fundamental
problem of quantum gravity looks rather uncertain. This concerns
both the method of Ref. \cite{ReuterWeyer} and the one of the old
works \cite{Oldpapers} where the perturbative higher derivative
quantum gravity effects were considered (and similar results were
obtained). {\it Second}, the present approach has at least the
virtue of introducing a parameter, $\nu$, whose definition
(namely, $\nu$ is the ratio of the masses squared of the heavy
matter fields versus the Planck mass squared, see
Eq.\,(\ref{nu})), naturally fits in the right order of magnitude
to explain the effect. At then end of the day, since both
approaches are essentially based on phenomenological input, the
preference of one of them remains a matter of taste. However, the
net outcome is that the running law (\ref{GH2}) and the one found
in \cite{ReuterWeyer} are identical if one identifies $q\equiv
2\,\nu$ in the limit of very small value of these parameters (the
only relevant limit for the galactic dynamics).

Let us also mention a recent modification of the RG approach, the
so-called metric-skew-tensor-gravity theory proposed in
\cite{Moffat}. Here a skew-symmetric field tensor
$F_{\mu\nu\lambda\sigma}$ is introduced and coupled to matter.
This model introduces a new particle, but the main contribution
to the mass of the galaxy is not from the skew field energy
density, but from the RG flow running of $G$ (induced by the
coupling of this field to ordinary matter), which leads to a
modified Newtonian acceleration law that can be successfully
fitted to the galaxy rotation curves. Again this is achieved for
special RG trajectories for which gravity becomes a (slowly
varying) ``confining force''. In this case the trajectory is
again the $q$-dependent one mentioned above, but for the original
value $q=2$. The price of returning to this value, which is
associated to a (hypothetical) IR fixed point, is to postulate the
existence of a new particle in the galaxy halo. The physical
consequence is again similar to our running law  for $G$, Eq.\,
(\ref{GH2}), which entails anti-screening for $\nu>0$, much in
the same way as the QCD force at small distances.

We find suggestive that all these RG approaches, even though they
differ in the particular details, seem to converge to the
same basic idea that the flat rotation curves of the galaxies
could perhaps be understood without using the hypothesis of CDM.
At the moment these RG frameworks cannot account for subleading
effects, such as the Tully-Fisher law. However, they may
constitute a promising framework to eventually construct the
theory of the flat rotation curves of the galaxies from the first
principles of Quantum Field Theory. The necessary modifications
to account for subleading effects can be a matter of improving the
technicalities of the RG framework and of the scale-setting
procedure, including a more accurate modeling of the galaxy.

\vspace{0.3cm}

{\bf Acknowledgments:}  The work of I.Sh. has been supported in
part by CNPq, FAPEMIG and ICTP. J.S. has been
supported in part by MECYT and FEDER under project FPA2001-3598.
J.S. is also thankful to the Max Planck Institut f\"ur Physik in
Munich for the warmth hospitality and the financial support
provided. The work of H. \v{S}tefan\v{c}i\'{c} is supported by
the Ministry of Science, Education and Sport of the Republic of
Croatia under the contract  No. 0098002.


\end{document}